\documentclass[pre,twocolumn]{revtex4-2}
\usepackage{amsmath}
\usepackage{amssymb}
\usepackage{graphicx}
\usepackage{braket}
\usepackage{stmaryrd}
\usepackage{hyperref}
\usepackage{color}
\usepackage{dsfont}
\usepackage{bm}

\hypersetup{
    colorlinks,
    citecolor=blue,
    linkcolor=blue,
    urlcolor=blue
}

\begin{document}

\title{
Fundamental Precision Limits in Finite-Dimensional Quantum Thermal Machines
}
\author{Yoshihiko Hasegawa}
\email{hasegawa@biom.t.u-tokyo.ac.jp}
\affiliation{Department of Information and Communication Engineering, Graduate
School of Information Science and Technology, The University of Tokyo,
Tokyo 113-8656, Japan}

\date{\today}
\begin{abstract}

Enhancing the precision of a thermodynamic process inevitably necessitates a thermodynamic cost.
This notion was recently formulated as the thermodynamic uncertainty relation, which states that
the lower bound on
the relative variance of thermodynamic currents decreases as entropy production increases. 
From another viewpoint,
the thermodynamic uncertainty relation implies that if entropy production were allowed to become infinitely large,  the lower bound on the relative variance could approach zero.
However, it is evident that realizing infinitely large entropy production is infeasible in reality. 
This indicates that physical constraints impose precision limits on the system, independent of its dynamics.
In this study, we derive  fundamental precision limits,  dynamics-independent bounds on the relative variance and the expectations of observables for open quantum thermal machines operating within a finite-dimensional system and environment.
These bounds are set by quantities such as dimensions and energy bandwidth, which depend only on the initial configuration and are independent of the dynamics.
Using a quantum battery model, 
the fundamental precision limits show
that there is a trade-off between the amount of energy storage and the charging precision.
Additionally, 
we investigate how quantum coherence affects these 
fundamental
limits, demonstrating that the presence of coherence can improve the precision limits. Our findings provide insights into fundamental limits on the precision of quantum thermal machines.

\end{abstract}
\maketitle

\begin{figure}
\includegraphics[width=1.0\linewidth]{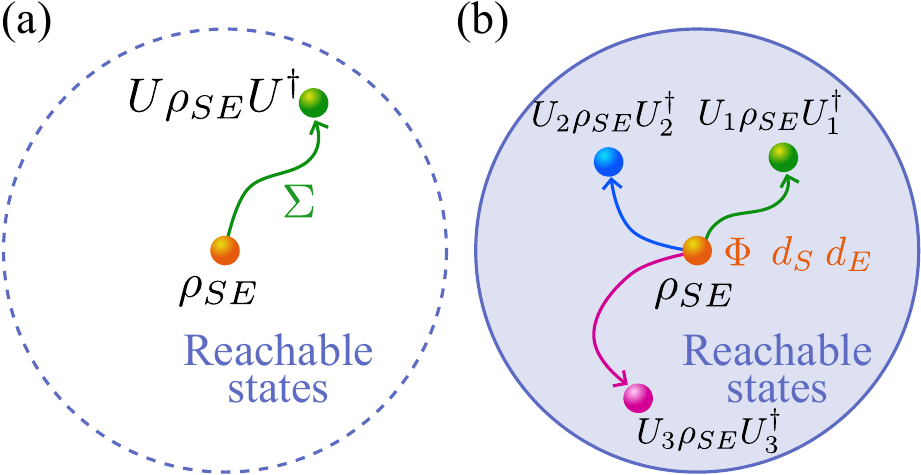}
\caption{
Conceptual illustration of thermodynamic bounds. 
Consider the initial state $\rho_{SE} = \rho_S \otimes \rho_E$, where the system undergoes a unitary transformation. The large circles represent the set of all states accessible from $\rho_{SE}$ by any possible unitary transformation.
(a) Thermodynamic uncertainty relations focus on a single time evolution under a specific unitary $U$. The precision is evaluated for the state $U \rho_{SE} U^\dagger$, and its lower bound is determined by the entropy production $\Sigma$, which depends on the chosen unitary $U$. As a result, this bound applies only to the particular time evolution generated by $U$.
(b) Fundamental precision limits, on the other hand, consider all possible time evolutions starting from $\rho_{SE}$, that is, any state within the circle. Here, the lower bound of the precision is determined by quantities that are independent of the specific unitary transformation. Therefore, these bounds apply universally to all states within the circle.
}
\label{fig:ponch}
\end{figure}

\textit{Introduction.---}In the physical world, there is no free lunch. This fundamental principle is formulated by the thermodynamic uncertainty relation, which shows that making a physical process more precise comes at a thermodynamic cost. 
It was first studied in classical stochastic thermodynamics \cite{Barato:2015:UncRel,Gingrich:2016:TUP,Garrahan:2017:TUR,Dechant:2018:TUR,Terlizzi:2019:KUR,Hasegawa:2019:CRI,Hasegawa:2019:FTUR,Dechant:2020:FRIPNAS,Vo:2020:TURCSLPRE,Koyuk:2020:TUR} within the framework of Markov processes. 
Consider a classical Markov process, where $C$ denotes a thermodynamic current over the time interval $[0,\tau]$, and $\tilde{\Sigma}$ denotes the entropy production during the interval $[0,\tau]$. The thermodynamic uncertainty relation indicates that the following lower bound exists for the relative variance of $C$ \cite{Barato:2015:UncRel,Gingrich:2016:TUP}:
\begin{align}
    \frac{\mathrm{Var}[C]}{\mathbb{E}[C]^{2}}\ge\frac{2}{\tilde{\Sigma}},
    \label{eq:conventional_TUR}
\end{align}
where $\mathbb{E}[C]$ and $\mathrm{Var}[C]$ denote the expectation and variance of $C$, respectively. 
Similar inequalities also hold for thermodynamic costs other than entropy production.
Dynamical activity $\tilde{A}$ within the interval $[0,\tau]$ is a physical quantity that quantifies the degree of activity in classical Markov processes,
leading to the inequality $\mathrm{Var}[C]/\mathbb{E}[C]^{2}\ge1/\tilde{A}$ \cite{Garrahan:2017:TUR,Terlizzi:2019:KUR}.
Particularly for dynamics far from equilibrium, the inequality derived from dynamical activity provides a tighter bound.
Equation~\eqref{eq:conventional_TUR} demonstrates that increasing the entropy production $\tilde{\Sigma}$ is necessary to improve the accuracy of a thermodynamic machine. 
Conversely, Eq.~\eqref{eq:conventional_TUR} indicates that if the entropy production could be increased without bound,
the lower bound of
the relative variance could be reduced to zero. 
However, in reality, it is impossible to generate an infinite amount of entropy production due to physical constraints. 
This issue becomes clearer when considering finite-dimensional quantum systems. 
The thermodynamic uncertainty relation in quantum thermodynamics \cite{Binder:2018:QuantumThermoBook,Potts:2024:QuantumThermodynamics} has been a focus of active research in recent years \cite{Erker:2017:QClockTUR,Brandner:2018:Transport,Carollo:2019:QuantumLDP,Liu:2019:QTUR,Guarnieri:2019:QTURPRR,Saryal:2019:TUR,Hasegawa:2020:QTURPRL,Hasegawa:2020:TUROQS,Kalaee:2021:QTURPRE,Monnai:2022:QTUR,Hasegawa:2023:BulkBoundaryBoundNC}. 
Previous studies \cite{Hasegawa:2020:QTURPRL,Vu:2021:QTURPRL,
Hasegawa:2023:BulkBoundaryBoundNC,
Nishiyama:2023:ExactQDA,Prech:2025:CoherenceQTUR,Vu:2025:Fundamental} have shown that the thermodynamic cost term or the expectation value terms incorporate quantum corrections that arise from the coherent dynamics, leading to enhanced precision in the presence of quantum coherence. 
Open quantum dynamics can be described by a joint unitary evolution of the principal system and the environment with finite dimensions.
For example, Ref.~\cite{Salazar:2024:QRelEntTUR} derived a quantum thermodynamic uncertainty relation comprising the system and the environment. 
A similar bound can be obtained by applying the monotonicity relation to a lower bound on the information divergence (cf. Eq.~\eqref{eq:QTUR_Hellinger}).
These thermodynamic uncertainty relations do not impose any limitations on the possible values of the quantum entropy production as long as it is non-negative. 
However, it is clear that infinite entropy production is not possible due to physical constraints; 
no matter how the system evolves over time, there is an upper bound to the amount of entropy production.
This observation shows that the precision is fundamentally limited by physical constraints, regardless of the time evolution.
This study establishes the fundamental precision limits for open quantum thermal machines operating within a finite-dimensional system and environment. We derive a lower bound for the expectation value ratio of the observable (cf. Eq.~\eqref{eq:gammaE_pnorm_TUR}). 
Considering a specific case of the derived bound, we can obtain a lower bound for the relative variance (cf. Eq.~\eqref{eq:gammaE_conv_TUR}) that mirrors the form of the thermodynamic uncertainty relation. Additionally, we obtain an upper bound for the expectation of the observable (cf. Eq.~\eqref{eq:gammaE_expectation_bound}). 
These bounds depend on factors such as dimensions, energy bandwidth, and initial minimum eigenvalue, which are determined solely by the initial configuration rather than by specific dynamics.
Therefore, the fundamental precision limits apply to any unitary operator acting on the composite system (Fig.~\ref{fig:ponch}). 
Using a quantum battery model as an example, we show that there is a trade-off between the amount of energy that can be stored and the charging precision.
Moreover, we show that the presence of coherence in the initial state can potentially improve the fundamental limit of the precision of quantum thermal machines.

\textit{
Fundamental
precision limits.---}Trade-offs between precision and cost have been traditionally studied in classical and quantum Markov processes.
Here, we consider a general open quantum dynamics, where the
 total 
system comprises the principal system and the environment. 
In principle, any Markov process falls into this category. 
This kind of setting was studied in Refs.~\cite{Hasegawa:2020:TUROQS,Salazar:2024:QRelEntTUR,Ishida:2024:QTURVerification} for deriving quantum thermodynamic uncertainty relations. 
To state clearly, we derive the precision bounds under the following physical constraints:
\begin{enumerate}
    \item The principal system and the environment, both of finite dimensions, are assumed to undergo a joint unitary evolution
\item The preparation of pure states is impossible due to the third law of thermodynamics \cite{Allahverdyan:2011:Cooling, Reed:2014:Landauer, Clivaz:2019:Cooling, Buffoni:2022:ThirdLawQC,Buffoni:2022:ThirdLawQC}
\end{enumerate}

\begin{figure}
\includegraphics[width=0.95\linewidth]{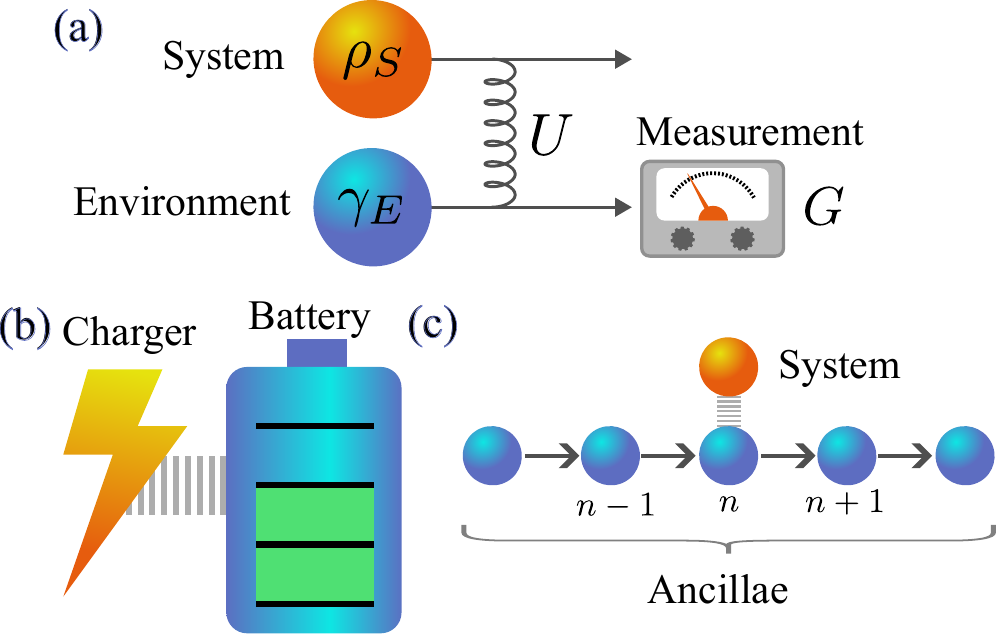}    
\caption{
Open quantum models. 
(a)
The basic model comprising the system $S$ and the environment $E$.
The joint unitary $U$ acts on $S$ and $E$; after the interaction, $E$ is measured with a Hermitian observable $G$.
Here, we assume that the initial environmental state is the Gibbs state $\gamma_E$. 
(b) In the quantum battery model, the battery is charged through its interaction with a charger. If we regard the charger as the system and the battery as the environment, this charging process falls within the framework of (a).
(c) In the collision model, the system interacts sequentially with ancillae, and each ancilla is measured after its interaction. If we treat the ancillae as the environment, this model also falls within the framework of (a).
}
\label{fig:model_description}
\end{figure}

Consider an open quantum thermal machine consisting of the system $S$ and the environment $E$ (Fig.~\ref{fig:model_description}(a)). 
We can consider the state transformation of the composite system by applying a joint unitary operator $U$.
Let $H_S$ and $H_E$ be the Hamiltonian operators representing the system $S$ and the environment $E$, respectively. These operators can be expressed in their spectral decompositions as follows:
\begin{align}
    H_{S}=\sum_{i=1}^{d_{S}}\sigma_{i}\ket{\sigma_{i}}\bra{\sigma_{i}},\quad H_{E}=\sum_{i=1}^{d_{E}}\epsilon_{i}\ket{\epsilon_{i}}\bra{\epsilon_{i}}.\label{eq:HSHE_def}
\end{align}
Here, $\sigma_i$ and $\epsilon_i$ denote the eigenvalues of $H_S$ and $H_E$, respectively, and $\ket{\sigma_i}$ and $\ket{\epsilon_i}$ represent the corresponding eigenvectors. The dimensions of the system and the environment are given by $d_S$ and $d_E$, respectively.
After preparing the initial states, we apply the unitary transformation $U$ to the composite system:
\begin{align}
    \rho_{SE}^{\prime}=U(\rho_{S}\otimes\rho_{E})U^{\dagger},
    \label{eq:rho_SE_joint_unitary}
\end{align}
where $\rho_S$ and $\rho_E$ are the system and environment initial states, respectively, and $\rho_{SE}^\prime$ is the joint state after the time evolution. 
Here, we have assumed that the initial state is a product state.
Note that it is impossible to prepare states with vanishing eigenvalues. If such a state could be prepared, it would allow for the creation of pure states, which contradicts the third law of thermodynamics.

Let $\rho_E^\prime \equiv \mathrm{Tr}_S[\rho_{SE}^\prime]$ be the state of the environment after the unitary operation $U$. 
We wish to measure the environment after the unitary evolution (Fig.~\ref{fig:model_description}(a)). 
This approach is known as indirect measurement, where rather than directly measuring the system itself, an ancillary system is used to measure the system. By measuring the ancillary system, information about the principal system can be inferred.
For example, the quantum battery model \cite{Campaioli:2018:QuantumBatteries,Campaioli:2024:QuantumBatteries} and the collision model \cite{Ciccarello:2022:Collision} belong to the class of the joint-unitary evolution model. In the quantum battery model, a battery stores energy by interacting unitarily with a charger that is initially charged (Fig.~\ref{fig:model_description}(b)). In the collision model (Fig.~\ref{fig:model_description}(c)), the environment is represented as a stream of ancillae that interact with the system sequentially for short durations. This model forms the basis for the continuous measurement formalism \cite{Landi:2023:CurFlucReviewPRXQ}, which is widely used in studies of quantum thermodynamic uncertainty relations \cite{Hasegawa:2020:QTURPRL}.

Let $G$ be an arbitrary observable measured on the environment $E$.
Let $\mathbb{E}_\rho[G] \equiv \mathrm{Tr}[\rho G]$ denote the expectation of $G$ with respect to $\rho$ and $\mathrm{Var}_\rho[G] \equiv \mathbb{E}_\rho[G^2] - \mathbb{E}_\rho[G]^2$ denote the variance of $G$. 
Given the observable $G$, we want to obtain a lower bound on $\mathrm{Var}_{\rho_{E}^{\prime}}[G]/\mathbb{E}_{\rho_{E}^{\prime}}[G]^{2}$,
which is the relative variance evaluated for the environment after the unitary transformation.
Let $\lambda(G)$ be a set of eigenvalues of $G$, and $\lambda_{\min}(G) \equiv \min \lambda(G)$ and $\lambda_{\max}(G) \equiv \max \lambda(G)$. 
Although we aim to explore general observables, it is necessary to impose a condition on $G$:
the minimum eigenvalue of $G$ is $0$, $\lambda_{\min}(G)=0$. 
Without this condition, we can make the relative variance arbitrarily small by transforming $G$ to $G + a\mathbb{I}$, where $a$ is a real value, and letting $a$ approach infinity. 
This transformation allows the mean value of $G$ to become infinitely large while maintaining a constant variance, $\mathrm{Var}_{\rho_E^\prime}[G]$. Consequently, the relative variance $\mathrm{Var}_{\rho_E^\prime}[G]/\mathbb{E}_{\rho_E^\prime}[G]^2$ can be made infinitely small, which is undesirable for our analysis.

Consider the ratio $\mathbb{E}_{\rho_E^\prime}\left[G^{s}\right]^{r/(s-r)}/\mathbb{E}_{\rho_E^\prime}\left[G^{r}\right]^{s/(s-r)}$, where $0<r<s$. 
For $r=1$ and $s=2$, $\mathbb{E}_{\rho_E^\prime}\left[G^{s}\right]^{r/(s-r)}/\mathbb{E}_{\rho_E^\prime}\left[G^{r}\right]^{s/(s-r)}-1$ becomes the relative variance, $\mathrm{Var}_{\rho_E^\prime}[G]/\mathbb{E}_{\rho_E^\prime}[G]^2$. 
Using the Petrov inequality \cite{Valentin:2007:TailProb}, the following relation holds:
\begin{align}
    \frac{\mathbb{E}_{\rho_{E}^{\prime}}\left[G^{s}\right]^{r/(s-r)}}{\mathbb{E}_{\rho_{E}^{\prime}}\left[G^{r}\right]^{s/(s-r)}}\ge\frac{1}{1-\lambda_{\min}(\rho_{E}^{\prime})},
    \label{eq:pzero_norm_TUR}
\end{align}
whose derivation is shown in the End Matter. 
According to Eq.~\eqref{eq:pzero_norm_TUR}, minimizing the relative variance of the observable $G$ requires minimizing the smallest eigenvalue, denoted as $\lambda_{\min}(\rho_E^\prime)$. 
When there is degeneracy in $G$, $\lambda_{\min}(\rho_E^\prime)$ should be replaced by $\lambda_{\min}(\rho_E^\prime)\delta_0$,
where $\delta_0$ represents
the multiplicity of the zero eigenvalue of $G$ (see the Supplementary Material \cite{Supp:2025:FundamentalUR}).
\nocite{Shiraishi:2018:SpeedLimit,Nishiyama:2023:EPUpperBound,MullerLennert:2013:RenyiEntropy,Holevo:1973:CRI,Dashti:2017:Bayes,Kosloff:2013:QThermoReview,VandenBroeck:2015:Review,Alicki:2013:QB,Scarani:2002:ThermalizingCollisionModel,Santos:2020:JFT,Gross:2018:ContMeasQubit,Proesmans:2017:TUR}
Given that $\delta_0 \ge 1$, Eq.~\eqref{eq:pzero_norm_TUR} must apply even in the degenerate scenario. Consequently, the subsequent analyses are also applicable to the degenerate situation.
Note that we recently derived a thermodynamic uncertainty relation using the Petrov inequality, where the bound comprises the dynamical activity \cite{Hasegawa:2024:ConcentrationIneqPRL}.
The problem of minimizing the smallest eigenvalue is equivalent to the problem of thermodynamic cooling \cite{Allahverdyan:2011:Cooling,Reed:2014:Landauer,Scharlau:2018:QuantumHorn,Clivaz:2019:Cooling,Oftelie:2024:DynamicCooling},
where ancillary systems are used to cool the target system. 
Then, we obtain (see the End Matter)
\begin{align}
    \frac{\mathbb{E}_{\rho_{E}^{\prime}}\left[G^{s}\right]^{r/(s-r)}}{\mathbb{E}_{\rho_{E}^{\prime}}\left[G^{r}\right]^{s/(s-r)}}\ge\frac{1}{1-d_{S}\lambda_{\min}(\rho_{S})\lambda_{\min}(\rho_{E})}.
    \label{eq:pzero_norm_TUR3}
\end{align}
From the assumption, $\lambda_{\min}(\rho_S)>0$ and $\lambda_{\min}(\rho_E)>0$.
Let $\gamma_E$ be the Gibbs state:
\begin{align}
    \gamma_{E}\equiv\frac{e^{-\beta H_{E}}}{Z_{E}(\beta)},
    \label{eq:gamma_E_def}
\end{align}
where $\beta$ is the inverse temperature and $Z_E(\beta) \equiv \mathrm{Tr}[e^{-\beta H_E}]$ is the partition function. 
Up to this point, no assumptions have been made regarding the initial states $\rho_S$ and $\rho_E$. From now on, the initial state of the environment will be taken as the Gibbs state, $\rho_E = \gamma_E$.
This scenario is often considered in open quantum systems \cite{Esposito:2010:EntProd,Reed:2014:Landauer}.
In this scenario, $\lambda_{\min}(\gamma_E)$ is bounded from below by \cite{Reed:2014:Landauer}
\begin{align}
    \lambda_{\min}\left(\gamma_{E}\right)=\frac{e^{-\beta\epsilon_{\max}}}{\mathrm{Tr}\left[e^{-\beta H_{E}}\right]}\ge\frac{e^{-\beta\epsilon_{\max}}}{d_{E}e^{-\beta\epsilon_{\min}}},
    \label{eq:P_min_eigenvalue}
\end{align}
where $\epsilon_{\max} \equiv \lambda_{\max}(H_E)$ and $\epsilon_{\min}\equiv \lambda_{\min}(H_E)$. 
From Eqs.~\eqref{eq:pzero_norm_TUR3} and \eqref{eq:P_min_eigenvalue},
we obtain
\begin{align}
\frac{\mathbb{E}_{\rho_{E}^{\prime}}\left[G^{s}\right]^{r/(s-r)}}{\mathbb{E}_{\rho_{E}^{\prime}}\left[G^{r}\right]^{s/(s-r)}}\ge\left(1-\frac{d_{S}}{d_{E}}e^{-\Phi(\beta,\Delta\epsilon,\lambda_{\min}(\rho_{S}))}\right)^{-1},
    \label{eq:gammaE_pnorm_TUR}
\end{align}
where $\Delta \epsilon \equiv \epsilon_{\max} - \epsilon_{\min}$ is the energy bandwidth of the environment $E$ and we define
\begin{align}
\Phi(\beta,\Delta\epsilon,\lambda_{\min}(\rho_{S}))\equiv\ln\frac{1}{\lambda_{\min}(\rho_{S})}+\beta\Delta\epsilon.
    \label{eq:cost_Phi_def}
\end{align}
For $s=2$ and $r=1$, Eq.~\eqref{eq:gammaE_pnorm_TUR} provides the lower bound to the relative variance:
\begin{align}
\frac{\mathrm{Var}_{\rho_{E}^{\prime}}[G]}{\mathbb{E}_{\rho_{E}^{\prime}}\left[G\right]^{2}}\ge\left(\frac{d_{E}}{d_{S}}e^{\Phi(\beta,\Delta\epsilon,\lambda_{\min}(\rho_{S}))}-1\right)^{-1}.
    \label{eq:gammaE_conv_TUR}
\end{align}
Equation~\eqref{eq:gammaE_conv_TUR} shows that, to make the lower bound of the relative variance smaller, the energy bandwidth $\Delta \epsilon$, quantified by the difference between the largest and smallest eigenvalues of the environment, should be sufficiently large. 
In addition, the dimensions, $d_S$ and $d_E$, and the smallest eigenvalue $\lambda_{\min}(\rho_S)$ play an important role in the bound;
a larger environment and smaller $\lambda_{\min}(\rho_S)$ ensure a more accurate thermal machine.
As can be seen, the bound depends on properties of the environment, which are often inaccessible.
In recent years, there have been studies of open systems using small, finite-sized environments. For example, Ref.~\cite{Cech:2023:QTrajQC} used a transmon quantum computer for simulating open quantum dynamics. In such systems, the energy gap is approximately $5$GHz and the temperature is $15$mK \cite{Koch:2007:Qubit,Gambetta:2017:Qubit}, making calculations of the bounds in Eqs.~\eqref{eq:gammaE_pnorm_TUR} and \eqref{eq:gammaE_conv_TUR} feasible.
Moreover, taking $s=\infty$ and $r=1$, we have
\begin{align}
\mathbb{E}_{\rho_{E}^{\prime}}\left[G\right]\le\lambda_{\max}[G]\left\{ 1-\frac{d_{S}}{d_{E}}e^{-\Phi(\beta,\Delta\epsilon,\lambda_{\min}(\rho_{S}))}\right\} .
    \label{eq:gammaE_expectation_bound}
\end{align}
This bound is reminiscent of the bound derived in Refs.~\cite{Hasegawa:2024:TCI,Hasegawa:2024:ConcentrationIneqPRL}. 
Equations~\eqref{eq:gammaE_pnorm_TUR}, \eqref{eq:gammaE_conv_TUR}, and \eqref{eq:gammaE_expectation_bound} are referred to as \textit{fundamental precision limits} and are the main results of this Letter. 
The lower bound in the conventional thermodynamic uncertainty relation comprises the entropy production, which depends on the actual time evolution of the system. 
In other words, the entropy production depends on the dynamics. 
However, the bounds in Eqs.~\eqref{eq:gammaE_pnorm_TUR}, \eqref{eq:gammaE_conv_TUR}, and \eqref{eq:gammaE_expectation_bound} depend solely on the initial configuration and do not involve quantities related to the time evolution.
This implies that Eqs.~\eqref{eq:gammaE_pnorm_TUR}, \eqref{eq:gammaE_conv_TUR}, and \eqref{eq:gammaE_expectation_bound} should hold for any conceivable joint unitary applied to the system and environment (Fig.~\ref{fig:ponch}). 
In this respect, the fundamental precision limits can be said to be more foundational than the thermodynamic uncertainty relation (Fig.~\ref{fig:ponch}).
In Eqs.~\eqref{eq:gammaE_pnorm_TUR}, \eqref{eq:gammaE_conv_TUR} and \eqref{eq:gammaE_expectation_bound}, we have assumed that the minimum eigenvalue of $G$ is zero. 
We can easily generalize the bounds to observables that do not satisfy this condition by replacing $G$ with $G - \lambda_{\mathrm{min}}(G)\mathbb{I}$ therein. 

\textit{Example.}---
As an application of the fundamental precision limits, we consider the quantum battery model \cite{Campaioli:2018:QuantumBatteries,Campaioli:2024:QuantumBatteries}. 
We consider a charger-mediated protocol in which the charger initially stores energy and transfers it to the battery through their interaction (Fig.~\ref{fig:model_description}(b)). 
When we identify the charger and the battery as $S$ and $E$, respectively,
the charging process falls into the model described by Eq.~\eqref{eq:rho_SE_joint_unitary}. 
In this case,
the battery Hamiltonian is $H_E$ defined in Eq.~\eqref{eq:HSHE_def}, and $\rho_E$ and $\rho_E^\prime$ denote the battery's initial and post-charging states, respectively. 
A primary performance metric is the stored energy $\mathcal{E}_{B}\equiv\mathbb{E}_{\rho_{E}^{\prime}}[H_{E}]-\mathbb{E}_{\rho_{E}}[H_{E}]$. 
Studies \cite{Friis:2018:PrecisionQB,Bakhshinezhad:2024:TradeoffsQB} used the post-charging energy variance, the charging precision, as another metric for quantum batteries, which is expressed as $\mathcal{P}_{B}\equiv\mathrm{Var}_{\rho_{E}^{\prime}}[H_{E}]$.
Therefore, an ideal quantum battery should have high energy storage $\mathcal{E}_B$ while keeping charging precision $\mathcal{P}_B$
 as small as possible 
\cite{Friis:2018:PrecisionQB,Bakhshinezhad:2024:TradeoffsQB}. It is natural to minimize the quantity $\mathcal{P}_B/\mathcal{E}_B^{2}$,
which reflects smaller variance and larger energy storage simultaneously. 
Using Eq.~\eqref{eq:gammaE_conv_TUR}, we obtain 
\begin{align}
    \frac{\mathcal{P}_{B}}{\mathcal{E}_{B}^{2}}\ge\left(\frac{d_{E}}{d_{S}}e^{\Phi(\beta,\Delta\epsilon,\lambda_{\min}(\rho_{S}))}-1\right)^{-1},
    \label{eq:EP_tradeoff}
\end{align}
which establishes a fundamental trade-off between stored energy and precision (see the Supplementary Material \cite{Supp:2025:FundamentalUR} for details).
In the Supplementary Material \cite{Supp:2025:FundamentalUR}, we consider the collision model (Fig.~\ref{fig:model_description}(c)) for another application.

\textit{Relation to entropy production.---}As mentioned in the Introduction, the quantum thermodynamic uncertainty relation has a lower bound comprising the entropy production $\Sigma$. 
$\Phi$ as defined in Eq.~\eqref{eq:cost_Phi_def} has the term $\beta \Delta \epsilon$ and it is naturally expected that larger $\beta \Delta \epsilon$  leads to a larger upper bound on  entropy production. 
Recently, Ref.~\cite{Vu:2024:3rdTradeOff} established a lower bound for the entropy production based on the minimum eigenvalue. 
We explore how  $\Phi$ defined in Eq.~\eqref{eq:cost_Phi_def}  serves as an upper bound for entropy production.
As before, suppose that the initial state of the environment is the Gibbs state $\rho_E=\gamma_E$. 
The quantum entropy production is given by \cite{Esposito:2010:EntProd,Reed:2014:Landauer}
\begin{align}
    \Sigma=\mathcal{S}(\rho_{S}^{\prime})-\mathcal{S}(\rho_{S})+\beta\left(\mathbb{E}_{\rho_{E}^{\prime}}[H_{E}]-\mathbb{E}_{\rho_{E}}[H_{E}]\right),
    \label{eq:EP_decomposition}
\end{align}
where  $\mathcal{S}(\rho)\equiv-\mathrm{Tr}\left[\rho\ln\rho\right]$  is the von Neumann entropy. 
The von Neumann entropy is bounded by $0 \le \mathcal{S}(\rho_S)\le \ln d_S$ for an arbitrary $\rho_S$. The expectation value satisfies
$\epsilon_{\min}\le\mathrm{Tr}[H_{E}\rho_{E}]\le\epsilon_{\max}$ for an arbitrary $\rho_E$. Combining these bounds, 
it can be shown that
\begin{align}
\Sigma\le\ln d_{S}+\beta\Delta\epsilon=\Phi(\beta,\Delta\epsilon,1/d_{S}).
    \label{eq:Sigma_upperbound}
\end{align}
Equation~\eqref{eq:Sigma_upperbound} shows that $\Phi$ is indeed related to the entropy production $\Sigma$; $\Phi$ provides an upper bound on $\Sigma$
when we replace $\lambda_{\min}(\rho_S)$ with $1/d_S$.
Moreover, 
since $\lambda_{\min}(\rho_S)\le 1/d_S$,
$\Phi(\beta,\Delta\epsilon,1/d_{S})\le\Phi(\beta,\Delta\epsilon,\lambda_{\min}(\rho_{S}))$.

Reference~\cite{Salazar:2024:QRelEntTUR} derived a quantum thermodynamic uncertainty relation for joint unitary evolution, whose lower bound comprises the entropy production $\Sigma$ in Eq.~\eqref{eq:EP_decomposition}. However, this relation includes the conjugate entropy production, which lacks a clear physical interpretation. By applying a lower bound on the information divergence \cite{Nishiyama:2020:HellingerBound}, we can derive an alternative thermodynamic uncertainty relation for joint unitary evolution that involves only the standard entropy production. This approach yields the following relation (see the Supplementary Material \cite{Supp:2025:FundamentalUR} for details):
\begin{align}
    \left(\frac{\sqrt{\mathrm{Var}_{\rho_{E}^{\prime}}[G]}+\sqrt{\mathrm{Var}_{\gamma_{E}}[G]}}{\mathbb{E}_{\rho_{E}^{\prime}}[G]-\mathbb{E}_{\gamma_{E}}[G]}\right)^{2}\ge\frac{1}{e^{\Sigma}-1}.
    \label{eq:QTUR_Hellinger}
\end{align}
By substituting the upper bound of the entropy production shown in
Eq.~\eqref{eq:Sigma_upperbound} into Eq.~\eqref{eq:QTUR_Hellinger}, we obtain
\begin{align}
    \left(\frac{\sqrt{\mathrm{Var}_{\rho_{E}^{\prime}}[G]}+\sqrt{\mathrm{Var}_{\gamma_{E}}[G]}}{\mathbb{E}_{\rho_{E}^{\prime}}[G]-\mathbb{E}_{\gamma_{E}}[G]}\right)^{2}\geq\frac{1}{e^{\Phi(\beta,\Delta\epsilon,1/d_{S})}-1},
    \label{eq:QTUR_Phi}
\end{align}
where $G$ is an arbitrary observable in the environment. 
Similarly to Eq.~\eqref{eq:gammaE_conv_TUR}, Eq.~\eqref{eq:QTUR_Phi} is a relation that holds for any unitary operator. 
However, the left side of Eq.~\eqref{eq:QTUR_Phi} represents the relative variance of $G$ concerning the time-evolved state $\rho_E^\prime$ and the initial state $\gamma_E$,
whose physical interpretation is not immediately clear.
This contrasts with Eq.~\eqref{eq:gammaE_conv_TUR}, where the left-hand side is the relative variance in the final state.
We provide a detailed comparison in 
the Supplementary Material \cite{Supp:2025:FundamentalUR}. 
Moreover, in 
the Supplementary Material, we also indicate a relation between the 
bounds, given in Eqs.~\eqref{eq:gammaE_pnorm_TUR}, \eqref{eq:gammaE_conv_TUR}, and \eqref{eq:gammaE_expectation_bound},
and the dynamical activity in classical Markov processes.

\textit{Coherent initial state.---}We next consider the case where there is quantum coherence in the initial environmental state. 
Quantum coherence is essential in numerous quantum technologies, enhancing their performance and enabling capabilities beyond what is possible in classical systems \cite{Streltsov:2017:CoherenceReview}. For example, quantum coherence allows heat engines to achieve efficiencies beyond the Carnot limit \cite{Scully:2003:QuantumCoherence,Rossnagel:2014:HeatEngine}. 
References~\cite{Liu:2021:QuantumRefrigerators,Kalaee:2021:QTURPRE} showed that the existence of coherence improves the lower bounds on thermodynamic uncertainty relations. 
Here, we show that this is also the case for the fundamental precision limits. 
In the preceding section, we used the Gibbs state for the initial density operator of the environment $E$, which is diagonal with respect to the Hamiltonian $H_E$. 
Here, 
suppose that the initial density operator of the environment is given by $\rho_E^c$, where
\begin{align}
    \rho_E^c \equiv \gamma_E + \chi_E.
    \label{eq:coherent_initial}
\end{align}
Here, $\chi_E$ is a Hermitian operator that has no diagonal elements when represented in the energy eigenbasis of $H_E$. 
The state $\rho_E^c$ appears frequently in the quantum
thermodynamics
literature as a tool for analyzing the role of quantum coherence in physical systems \cite{Rodrigues:2019:CoherentCollision,Hammam:2022:Coherence,Huang:2024:NonthermalReservoirs}. 
Consequently, the diagonal elements of $\rho_E^c$, which represent the population, follow the Gibbs distribution with respect to $H_E$.
Let us quantify the extent of coherence by the Frobenius norm of $\chi_E$:
\begin{align}
    \mathcal{C}\equiv\left\Vert \chi_{E}\right\Vert _{F}=\sqrt{\sum_{i,j}\left|\chi_{E,ij}\right|^{2}}.
    \label{eq:chi_Frobenius_main}
\end{align}
The Frobenius norm is used to measure the extent of quantum coherence, particularly since the diagonal elements of $\chi_E$ are zero. 
Thus, the coherence measure $\mathcal{C}$ quantifies the degree of quantum coherence, with $\mathcal{C} = 0$ indicating the absence of coherence.
We obtain the following bound:
\begin{align}
\frac{\mathbb{E}_{\rho_{E}^{\prime}}\left[G^{s}\right]^{r/(s-r)}}{\mathbb{E}_{\rho_{E}^{\prime}}\left[G^{r}\right]^{s/(s-r)}}\ge\left[1-\left(\frac{d_{S}}{d_{E}}e^{-\Phi}-d_{S}\lambda_{\min}(\rho_{S})\mathcal{C}\right)\right]^{-1},
    \label{eq:gammaE_pnorm_TUR_coherence}
\end{align}
which is valid when $(d_{S}/d_{E})e^{-\Phi}-d_{S}\lambda_{\min}(\rho_{S})\mathcal{C}$ is positive. 
Equation~\eqref{eq:gammaE_pnorm_TUR_coherence} shows the
 fundamental precision limit in the presence of coherence. 
Since the coherence term is $\mathcal{C}\ge 0$, the presence of coherence improves  the fundamental precision limits.

\textit{Conclusion.---}
We have derived fundamental precision limits for finite-dimensional quantum thermal machines that are independent of the specific dynamics. Our main result shows that the relative variance of observables is bounded from below by quantities that depend only on initial state properties, not on the particular unitary evolution. This represents a more fundamental constraint than conventional thermodynamic uncertainty relations.
From the perspective of experimental quantum systems, Eq.~\eqref{eq:gammaE_pnorm_TUR} plays a particularly important role. 
In practice, the initial state of the environment in quantum devices is typically mixed, which ensures that the smallest eigenvalue remains greater than zero. 
This constraint is crucial when testing thermodynamic uncertainty relations with actual quantum hardware. Recent studies have demonstrated growing interest in experimentally verifying such theoretical relations using quantum computers \cite{Gardas:2018:QuantumAnnealers,Solfanelli:2021:FluctuationRelations,Zhang:2022:EntropyProduction}. For example, in our previous work \cite{Ishida:2024:QTURVerification}, we experimentally confirmed a variant of the quantum thermodynamic uncertainty relation \cite{Hasegawa:2020:TUROQS} using the IBM quantum computer. 
Therefore, Eq.~\eqref{eq:gammaE_pnorm_TUR} provides a realistic and experimentally relevant bound on achievable precision, reflecting the unavoidable constraints imposed by the mixedness of the environment in quantum systems.

\begin{acknowledgments}
The author would like to thank Tomohiro Nishiyama
 for fruitful comments. 
This work was supported by JSPS KAKENHI Grant Number JP23K24915.
\end{acknowledgments}

\subsection*{DATA AVAILABILITY}

The program that produces Fig.~S3 in the supplementary material is openly available \cite{Data:2025:FundamentalUR}.

\appendix

\section*{End Matter}

\section{Derivation of Eq.~\eqref{eq:pzero_norm_TUR}}

We show the derivation of the  fundamental  precision limits.
Suppose that $\rho_E^\prime$ and $G$ admit the eigenvalue decompositions:
\begin{align}
\rho_{E}^{\prime}&=\sum_{n=1}^{d_{E}}r_{n}\ket{n}\bra{n},\label{eq:rho_decomp}\\G&=\sum_{n=1}^{d_{E}}g_{n}\ket{g_{n}}\bra{g_{n}},
\label{eq:G_decomp}
\end{align}
where $r_n \ge 0$, $\sum_n r_n = 1$, and
$g_n \ge 0$. 
Here, we exclude the possibility of degeneracy in $G$ for simplicity; however, this case can be addressed if necessary (see the Supplementary Material \cite{Supp:2025:FundamentalUR} for details).
When a measurement is performed using the operator $G$, one of the eigenvalues $\lambda(G)=\{g_1, g_2, \ldots, g_{d_E}\}$ is observed.
Since we have assumed that $\lambda_{\min}(G)=0$, we set $g_1=0$ without loss of generality. 
Let $X$ be a random variable. 
The Petrov inequality \cite{Valentin:2007:TailProb} is given by
\begin{align}
P(|X|>b) \geq \frac{\left(\mathbb{E}\left[|X|^r\right]-b^r\right)^{s /(s-r)}}{\mathbb{E}\left[|X|^s\right]^{r /(s-r)}},
\label{eq:Petrov_ineq_def}
\end{align}
where $0 < r < s$, $b \geq 0$, and the condition $b^r \leq \mathbb{E}\left[|X|^r\right]$ must be satisfied. 
From Eq.~\eqref{eq:Petrov_ineq_def},
by setting $b=0$,
the lower bound on the relative fluctuation of $G$ is given by
\begin{align}
    \frac{\mathbb{E}_{\rho_{E}^{\prime}}\left[G^{s}\right]^{r/(s-r)}}{\mathbb{E}_{\rho_{E}^{\prime}}\left[G^{r}\right]^{s/(s-r)}}\ge\frac{1}{1-P(G=0)},
    \label{eq:Petrov_ineq2_main}
\end{align}
where we used the property that $G$ is non-negative. 
Here, $0<r<s$ and $P(G=0)\equiv\mathrm{Tr}[\rho_{E}^{\prime}\ket{g_{1}}\bra{g_{1}}]$ is the probability of observing $G=g_1=0$ given $\rho_E^\prime$. 
For $r=1$ and $s=2$, Eq.~\eqref{eq:Petrov_ineq2_main} gives a lower bound for the relative variance:
\begin{align}
    \frac{\mathrm{Var}_{\rho_{E}^{\prime}}[G]}{\mathbb{E}_{\rho_{E}^{\prime}}[G]^{2}}\ge\left(\frac{1}{P(G=0)}-1\right)^{-1}.
    \label{eq:TUR_by_PGzero}
\end{align}
Equations~\eqref{eq:Petrov_ineq2_main} and \eqref{eq:TUR_by_PGzero} suggest that minimizing $P(G=0)$ is essential for achieving higher precision. 
We can express $P(G=0)=\sum_{n}r_{n}c_{n}$,
where $c_n$ is defined by $c_{n}\equiv|\braket{n|g_{1}}|^{2}$,
and satisfies $c_n \ge 0$ and $\sum_{n} c_{n} = 1$.
Then the minimization problem can be formulated as follows:
\begin{align}
\underset{\{c_{n}\}}{\text{minimize}}\quad&\sum_{n}r_{n}c_{n}\label{eq:minimize}\\\text{subject to}\quad&0\le c_{n}\le1,\sum_{n}c_{n}=1.\label{eq:subject_to}
\end{align}
This minimization problem can be solved with linear programming. 
In linear programming, the optimal solution lies at an extreme point of the feasible region. In this case, extreme points are characterized by having exactly one of the $c_n$ equal to $1$ and the others equal to $0$. The solution achieving the minimum is selected from these extreme points.
Let $r_1$ be the minimum of $\lambda(\rho_E^\prime)$,
$r_1 = \lambda_{\min}(\rho_E^\prime)$. 
Therefore, the minimum of $P(G=0)$ is achieved when $\ket{n=1}=\ket{g_{1}}$, which yields
\begin{align}
    \frac{\mathbb{E}_{\rho_{E}^{\prime}}\left[G^{s}\right]^{r/(s-r)}}{\mathbb{E}_{\rho_{E}^{\prime}}\left[G^{r}\right]^{s/(s-r)}}\ge\frac{1}{1-r_{1}}=\frac{1}{1-\lambda_{\min}(\rho_{E}^{\prime})}.
    \label{eq:pzero_norm_TUR_methods}
\end{align}
Equation~\eqref{eq:pzero_norm_TUR_methods} is Eq.~\eqref{eq:pzero_norm_TUR} in the main text. 

By setting  $r=1$ and $s=2$  in this inequality, we have
\begin{align}
    \frac{\mathrm{Var}_{\rho_{E}^{\prime}}\left[G\right]}{\mathbb{E}_{\rho_{E}^{\prime}}\left[G\right]^{2}}\ge\frac{\lambda_{\min}(\rho_{E}^{\prime})}{1-\lambda_{\min}(\rho_{E}^{\prime})}.
    \label{eq:Petrov_ineq_mean_var}
\end{align}
When taking $r=1$ and $s\to \infty$, we have
\begin{align}
    \left(1-\lambda_{\min}(\rho_{E}^{\prime})\right)\lambda_{\max}(G)\ge\mathbb{E}_{\rho_E^\prime}\left[G\right].
    \label{eq:Petrov_ineq_max}
\end{align}
Here, we used $\mathbb{E}_{\rho_E^\prime}[G^{s}]^{1/s}\overset{s\to\infty}{\longrightarrow}\lambda_{\max}(G)$. 

\section{Minimization of the smallest eigenvalue}

We consider minimizing the smallest eigenvalue of $\rho_E^\prime.$ This problem setting is equivalent to the problem of thermodynamic cooling \cite{Allahverdyan:2011:Cooling,Reed:2014:Landauer,Scharlau:2018:QuantumHorn,Clivaz:2019:Cooling,Oftelie:2024:DynamicCooling},
where ancillary systems are used to cool the target system. 
$\rho_E^\prime$ is calculated by
\begin{align}
    \rho_{E}^{\prime}=\sum_{n=1}^{d_{S}}\left(\bra{h_{n}}\otimes\mathbb{I}\right)\rho_{SE}^{\prime}\left(\ket{h_{n}}\otimes\mathbb{I}\right),
    \label{eq:rhoE_partial_trace}
\end{align}
where $\ket{h_n}$ is a basis in $S$.
Therefore, when considering all possible unitary transformations, the minimum achievable eigenvalue of $\rho_E^\prime$ is identical to the sum of the $d_S$ smallest elements of $\lambda(\rho_{SE}^\prime)$ \cite{Allahverdyan:2011:Cooling,Clivaz:2019:Cooling}.
Moreover, the eigenvalues of $\rho_{SE}^\prime$ are identical to the eigenvalues of $\rho_{SE}$, which is expressed by
\begin{align}
    \lambda(\rho_{SE}) = \{ab|a\in\lambda(\rho_{S}),b\in\lambda(\rho_{E})\}.
    \label{eq:Z_set_def}
\end{align}
For a vector $x \in \mathbb{R}^n$, let $x^\uparrow$ represent the vector formed by sorting the components of $x$ in non-decreasing order, such that $x_1^\uparrow \le x_2^\uparrow \le \cdots \le x_n^\uparrow$, where $x_i^\uparrow$ is the $i$-th element in this sorted sequence. Similarly, let $x^\downarrow$ denote the vector obtained by arranging the components of $x$ in non-increasing order, so that $x_1^\downarrow \ge x_2^\downarrow \ge \cdots \ge x_n^\downarrow$, where $x_i^\downarrow$ is the $i$-th element in this sequence.
Then the minimum eigenvalue is bounded from below by
\begin{align}
    \lambda_{\min}(\rho_{E}^{\prime})\ge\sum_{i=1}^{d_{S}}\lambda_{i}^{\uparrow}(\rho_{SE}).
    \label{eq:r0_expr}
\end{align}
Equation~\eqref{eq:r0_expr} states that a lower bound of the minimum eigenvalue can be expressed as the sum of the $d_S$ smallest eigenvalues of $\rho_{SE}$.
Then the $d_S$ sum part in Eq.~\eqref{eq:r0_expr} can be bounded from below by
\begin{align}
    \sum_{i=1}^{d_{S}}\lambda_{i}^{\uparrow}(\rho_{SE})&\ge d_{S}\lambda_{\min}(\rho_{SE})\nonumber\\&=d_{S}\lambda_{\min}(\rho_{S})\lambda_{\min}(\rho_{E}),
    \label{eq:ev_sum_lowerbound}
\end{align}
where we used the condition $\rho_{SE} = \rho_S \otimes \rho_E$. 
Combining Eq.~\eqref{eq:pzero_norm_TUR} with \eqref{eq:ev_sum_lowerbound}, we obtain Eq.~\eqref{eq:pzero_norm_TUR3}.

\section{Coherent initial state case}

In the main text, we explore the  fundamental  precision limits in the presence of initial coherence. For the reader's convenience, we first provide a brief review of the theorems necessary for the derivation.

\subsubsection*{Schur-Horn theorem}

Let $A$ be an $N\times N$ Hermitian matrix and let $\mathfrak{d}(A)$ be a set
of diagonal elements of $A$. 
The Schur-Horn theorem states that there exists a Hermitian matrix with
diagonal values $\mathfrak{d}_{1}^{\downarrow}(A),\ldots,\mathfrak{d}_{N}^{\downarrow}(A)$
in this order (the top left is $\mathfrak{d}_{1}^{\downarrow}(A)$
and so on) and eigenvalues $\lambda_{1}^{\downarrow}(A),\ldots,\lambda_{N}^{\downarrow}(A)$
if and only if
\begin{align}
    \sum_{i=1}^{n}\mathfrak{d}_{i}^{\downarrow}(A)\leq\sum_{i=1}^{n}\lambda_{i}^{\downarrow}(A)\quad n=1,\ldots,N-1,
    \label{eq:SchurHorn1}
\end{align}
and
\begin{align}
    \sum_{i=1}^{N}\mathfrak{d}_{i}^{\downarrow}(A)=\sum_{i=1}^{N}\lambda_{i}^{\downarrow}(A).
    \label{eq:SchurHorn2}
\end{align}
The second condition given by Eq.~\eqref{eq:SchurHorn2} is trivial, as the trace should be equal to the
sum of eigenvalues. 
From Eqs.~\eqref{eq:SchurHorn1} and \eqref{eq:SchurHorn2}, it is evident that 
\begin{align}
    \lambda_{\min}(A)\le\mathfrak{d}_{\min}(A),
    \label{eq:lmin_dmin}
\end{align}
where $\mathfrak{d}_{\min}(A)$ is the minimum of the diagonal elements of $A$. 

\subsubsection*{Weyl inequality}

The Weyl inequality relates the eigenvalues of two Hermitian matrices to the eigenvalues of their sum. Consider two $N \times N$ Hermitian matrices $A$ and $B$. 
The Weyl inequality states that each eigenvalue of the sum $A+B$ is bounded by the sum of a corresponding eigenvalue of $A$ and either the smallest or largest eigenvalue of $B$. More precisely, for every $1 \leq k \leq N$:
\begin{align}
\lambda_k^\downarrow(A) + \lambda_N^\downarrow(B) &\leq \lambda_k^\downarrow(A+B) \leq \lambda_k^\downarrow(A) + \lambda_1^\downarrow(B).
\label{eq:Weyl_inequality}
\end{align}
Then, suppose that the matrix $A$ is perturbed by a Hermitian matrix $\Delta A$. Then, from Eq.~\eqref{eq:Weyl_inequality}, we have
\begin{align}
    \left|\lambda_{k}^\downarrow(A+\Delta A) - \lambda_{k}^\downarrow(A)\right| \le\left\Vert \Delta A\right\Vert _{\mathrm{op}},
    \label{eq:lambda_perturbation}
\end{align}
where $\left\Vert \Delta A\right\Vert _{\mathrm{op}}$ denotes the operator norm, which equals $\left\Vert \Delta A\right\Vert _{\mathrm{op}} = \max |\lambda(\Delta A)|$. 
Let $\|A\|_F$ be the Frobenius norm:
\begin{align}
    \|A\|_{F}=\sqrt{\sum_{i=1}^{N}\sum_{j=1}^{N}\left|a_{ij}\right|^{2}}.
    \label{eq:Frobenius_norm_def}
\end{align}
It is known that $\|A\|_{\mathrm{op}}\le\|A\|_{F}$, which yields
\begin{align}
    \left|\lambda_{k}^\downarrow(A+\Delta A) - \lambda_{k}^\downarrow(A)\right|\le\left\Vert \Delta A\right\Vert _{\mathrm{op}}\le\left\Vert \Delta A\right\Vert _{F}.
    \label{eq:lambda_perturbation2}
\end{align}

\subsubsection*{Derivation of Eq.~\eqref{eq:gammaE_pnorm_TUR_coherence}}

Using Eq.~\eqref{eq:lmin_dmin}, it can be shown that
\begin{align}
    \lambda_{\min}(\rho_E^c) \le \lambda_{\min}(\gamma_E).
    \label{eq:lambda_rhoEc}
\end{align}
Equation~\eqref{eq:lambda_rhoEc} guarantees that the minimum eigenvalue becomes less than or equal to its original value in the presence of quantum coherence.
However, 
the Schur-Horn theorem provides only the qualitative effect on the minimum eigenvalue when coherence is present.
According to Eq.~\eqref{eq:lambda_perturbation2} the following relation holds:
\begin{align}
\left|\lambda_{\min}(\rho_{E}^{c})-\lambda_{\min}(\gamma_{E})\right|\le\mathcal{C},
    \label{eq:lambda_rhoE_chiE}
\end{align}
where $\mathcal{C}$ is defined in Eq.~\eqref{eq:chi_Frobenius_main}. 
Equation~\eqref{eq:lambda_rhoE_chiE} leads to
\begin{align}
    \lambda_{\min}(\gamma_{E})-\mathcal{C}\le\lambda_{\min}(\rho_{E}^{c}).
    \label{eq:lmin_Cq}
\end{align}
Equation~\eqref{eq:lmin_Cq} shows that
the smallest eigenvalue is reduced by $\mathcal{C}$.
Using Eq.~\eqref{eq:lmin_Cq}, we can derive Eq.~\eqref{eq:gammaE_pnorm_TUR_coherence} in the main text.

\end{document}

\title{Supplementary Material for\\ ``Fundamental Precision Limits in Finite-Dimensional Quantum Thermal Machines''}

\author{Yoshihiko Hasegawa}
\email{hasegawa@biom.t.u-tokyo.ac.jp}
\affiliation{Department of Information and Communication Engineering, Graduate
School of Information Science and Technology, The University of Tokyo,
Tokyo 113-8656, Japan}

\maketitle
This supplementary material describes the calculations introduced in the main text. The numbers of the equations and the figures are prefixed with S (e.g., Eq.~(S1) or Fig.~S1). Numbers without this prefix (e.g., Eq.~(1) or Fig.~1) refer to items in the main text.

\section{Degenerate case\label{sec:degenerate}}

In the main text, we have considered the rank-1 projector for $G$. 
Here, we consider the case where $G$ has degeneracy. 
Given the observable $G$ that has degeneracy,
we aim to find the minimum of $P(G=0)$. 
Both $\rho_E^\prime$ and $G$ have the following eigenvalue decomposition:
\begin{align}
    \rho_{E}^{\prime}&=\sum_{r}r\Upsilon_{r},\label{eq:rho_decomp_deg}\\
    G&=\sum_{g}g\Lambda_{g},\label{eq:G_decomp_deg}
\end{align}
where $\Upsilon_r$ and $\Lambda_g$ are projectors corresponding to $r$ and $g$, respectively. 
The probability of measuring $G=0$ is given by
\begin{align}
    P(G=0)=\mathrm{Tr}[\rho_{E}^{\prime}\Lambda_{0}]=\sum_{r}r\mathrm{Tr}[\Upsilon_{r}\Lambda_{0}]=\sum_{r}r\mathbbl{c}_{r},
    \label{eq:PG_zero_def}
\end{align}
where $\mathbbl{c}_r \equiv \mathrm{Tr}[\Upsilon_{r}\Lambda_{0}]$. 
Using the Cauchy-Schwarz inequality, we have
\begin{align}
    \mathrm{Tr}[\Upsilon_{r}\Lambda_{0}]^{2}\le\mathrm{Tr}[\Upsilon_{r}^{2}]\mathrm{Tr}[\Lambda_{0}^{2}]=\mathrm{Tr}[\Upsilon_{r}]\mathrm{Tr}[\Lambda_{0}]=\mathrm{rank}[\Upsilon_{r}]\mathrm{rank}[\Lambda_{0}].
    \label{eq:cr_upperbound}
\end{align}
Therefore, the following relation holds:
\begin{align}
    \mathbbl{c}_{r}\le\sqrt{\mathrm{rank}[\Upsilon_{r}]\mathrm{rank}[\Lambda_{0}]}.
    \label{eq:cr_upperbound2}
\end{align}
Analogous to the non-degenerate case, the minimization problem is formulated as follows:
\begin{align}
    \underset{\{\mathbbl{c}_{r}\}}{\text{minimize}}\quad&\sum_{r}r\mathbbl{c}_{r}\label{eq:minimize_deg}\\\text{subject to}\quad&0\le \mathbbl{c}_{r}\le\sqrt{\mathrm{rank}[\Upsilon_{r}]\mathrm{rank}[\Lambda_{0}]},\nonumber\\&\sum_{r}\mathbbl{c}_{r}=\mathrm{rank}[\Lambda_{0}].\label{eq:subject_to_deg}
\end{align}
The case $\mathrm{rank}[\Lambda_0] = 1$ is discussed in the main text, so here we consider $\mathrm{rank}[\Lambda_0] \ge 2$. Similarly to the rank one case, one might attempt to set a single $\mathbbl{c}_r$ equal to $\mathrm{rank}[\Lambda_0]$ and all other $\mathbbl{c}_r$ to zero,
which satisfies the second constraint in Eq.~\eqref{eq:subject_to_deg} and constitutes an extreme point. 
However, this choice does not satisfy the first constraint in Eq.~\eqref{eq:subject_to_deg}
when $\mathrm{rank}[\Upsilon_r] < \mathrm{rank}[\Lambda_0]$.
In this case, the probability $P(G=0)$ also receives contributions from eigenvalues beyond the smallest one, so the probability is not minimized.
Therefore, the minimum is achieved when $\mathrm{rank}[\Upsilon_r] \ge \mathrm{rank}[\Lambda_0]$,
in which we set a single $\mathbbl{c}_r$ equal to $\mathrm{rank}[\Lambda_0]$ and all other $\mathbbl{c}_r$ to zero.
Then we obtain
\begin{align}
    P(G=0)=\lambda_{\min}(\rho_{E}^{\prime})\delta_{0},
    \label{eq:PG_zero_final}
\end{align}
where $\delta_0$ is the degree of degeneracy of $g=0$ (that is, $\mathrm{rank}(\Lambda_0)$). 
Equation~\pzeroUnormUTUR{} in the main text becomes
\begin{align}
    \frac{\mathbb{E}_{\rho_{E}^{\prime}}\left[G^{s}\right]^{r/(s-r)}}{\mathbb{E}_{\rho_{E}^{\prime}}\left[G^{r}\right]^{s/(s-r)}}\ge\frac{1}{1-\delta_{0}\lambda_{\min}(\rho_{E}^{\prime})}.
    \label{eq:pzero_norm_TUR_deg}
\end{align}

\section{Relation to dynamical activity\label{sec:relation_to_DA}}

We show that
the term $\Phi$ in Eq.~\costUPhiUdef{}
serves a similar function to the dynamical activity in classical Markov processes, which plays a central role in trade-off relations \cite{Garrahan:2017:TUR,Shiraishi:2018:SpeedLimit,Terlizzi:2019:KUR}.

We provide definitions of entropy production and dynamical activity in Markov processes.
Let $W_{nm}$ be the transition rate from the $m$th state to the $n$th state in a classical Markov process and let $P_m(t)$ be the probability that the system is in $m$th state at time $t$.  
Given a Markov process $\dot{P}(t)=WP(t)$, where $W \equiv \{W_{nm}\}_{nm}$ and $P(t)=\{P_m(t)\}_m$ is the probability vector, the dynamical activity rate $\dot{\tilde{A}}$ and the entropy production rate $\dot{\tilde{\Sigma}}$ are defined by
\begin{align}
\dot{\tilde{\Sigma}}&\equiv\sum_{n<m}\left(P_{m}W_{nm}-P_{n}W_{mn}\right)\ln\frac{P_{m}W_{nm}}{P_{n}W_{mn}},\label{eq:EP_sigma_def}\\
\dot{\tilde{A}}&\equiv\sum_{n<m}\left(P_{m}W_{nm}+P_{n}W_{mn}\right).
\label{eq:DA_a_def}
\end{align}
For simplicity, we assume that the system is in the steady state. 
Let $N(\tau)$ be an observable of a stochastic trajectory in the time interval $[0,\tau]$, which is $N(\tau)=0$ when there is no jump in the trajectory. 
In Ref.~\cite{Hasegawa:2024:ConcentrationIneqPRL}, 
it was shown that the probability of $N(\tau) = 0$ and
the dynamical activity are related via
\begin{align}
    P(N(\tau)=0)\geq e^{-\tilde{A}}.
    \label{eq:PN_DA_supp}
\end{align}
Analogously, in this Letter, we have derived the lower bound on $P(G=0)$ as follows:
\begin{align}
P(G=0)\geq\frac{d_{S}}{d_{E}}e^{-\Phi(\beta,\Delta\epsilon,\lambda_{\min}(\rho_{S}))}.
    \label{eq:PN_Phi_supp}
\end{align}
From Eqs.~\eqref{eq:PN_DA_supp} and \eqref{eq:PN_Phi_supp},
the key point of the analogy is that both $e^{-\tilde{A}}$ and 
$(d_{S}/d_{E})e^{-\Phi}$
serve as quantities that determine lower bounds for the zero probabilities of observables. 

Moreover, there is another indication which demonstrates the analogy between the dynamical activity and $\Phi$ in Eq.~\costUPhiUdef{}. 
It has been revealed that the dynamical activity rate constitutes an upper bound on the entropy production rate \cite{Nishiyama:2023:EPUpperBound}:
\begin{align}
\dot{\tilde{\Sigma}}\le\kappa(R)\dot{\tilde{A}},
    \label{eq:Sigma_DA_classical_upperbound}
\end{align}
where $R$ is the maximum transition rate ratio $R\equiv \max_{n\neq m}W_{mn}/W_{nm}$ and $\kappa(R)\equiv(\ln R)(R-1)/(R+1)$.
The entropy production can become arbitrarily large if we allow the jump rate ratio to be infinitely large.
Equation~\eqref{eq:Sigma_DA_classical_upperbound} states that, when there is the upper bound on the ratio $W_{mn}/W_{nm}$, as specified by $R$, the entropy production rate $\dot{\tilde{\Sigma}}$ is bounded from above by the dynamical activity $\tilde{A}$.
This is reminiscent of Eq.~\SigmaUupperbound{}, showing that $\Phi$ serves as an upper bound on the entropy production. 
These analogies suggest that $\Phi$, as defined in Eq.~\costUPhiUdef{}, plays a role similar to that of the dynamical activity.

\section{Derivation of quantum thermodynamic uncertainty relation\label{sec:QTUR_Hellinger}}

We present the derivation of a quantum thermodynamic uncertainty relation given by Eq.~\QTURUHellinger{} in the main text,
whose lower bound comprises the entropy production $\Sigma$. 
For arbitrary density operators $\rho$ and $\sigma$, let $\mathrm{Fid}(\rho,\sigma)$ be the quantum fidelity:
\begin{align}
    \mathrm{Fid}(\rho,\sigma)\equiv\left(\mathrm{Tr}\sqrt{\sqrt{\rho}\sigma\sqrt{\rho}}\right)^{2},
    \label{eq:fidelity_def}
\end{align}
which quantifies the similarity between the two quantum states. 
Let $\mathcal{D}(\rho\|\sigma)$ be the quantum relative entropy:
\begin{align}
    \mathcal{D}(\rho\|\sigma)\equiv\mathrm{Tr}\left[\rho(\ln\rho-\ln\sigma)\right].
    \label{eq:QRE_def_supp}
\end{align}
According to Theorem 7 in Ref.~\cite{MullerLennert:2013:RenyiEntropy},
the following relation holds between the fidelity $\mathrm{Fid}(\rho,\sigma)$ and the quantum relative entropy $\mathcal{D}(\rho\|\sigma)$:
\begin{align}
    e^{-\mathcal{D}(\rho\|\sigma)}\le\mathrm{Fid}(\rho,\sigma).
    \label{eq:Dalpha_monotonic}
\end{align}
We next derive an upper bound on the quantum fidelity using the probabilities associated with measurements on the environment $E$. 
Let $\{\mathcal{M}_x\}_x$ be a positive operator-valued measure (POVM) acting on the environment, and
let $P_E(x)$ and $P_E^\prime(x)$ denote the probabilities of measuring $x$ with the POVM, $P_E(x) = \mathrm{Tr}[\mathcal{M}_x \gamma_E]$ and $P_E^\prime(x) = \mathrm{Tr}[\mathcal{M}_x \rho_E^\prime]$. 
Here, we assume that the measurement output $x$ of $\mathcal{M}_x$ is a real value. 
Moreover, the fidelity satisfies the following relation:
\begin{align}
    \sqrt{\mathrm{Fid}(\rho_{E}^{\prime},\gamma_{E})}&\le\sum_{x}\sqrt{P_{E}^{\prime}(x)P_{E}(x)}\nonumber\\&=1-\mathrm{Hel}^{2}\left(P_{E}^{\prime},P_{E}\right),
    \label{eq:fidelity_ineq}
\end{align}
where $\mathrm{Hel}^2(\cdot, \cdot)$ is the Hellinger distance:
\begin{align}
    \mathrm{Hel}^{2}\left(P_{1},P_{2}\right)\equiv\frac{1}{2}\sum_{x}\left(\sqrt{P_{1}(x)}-\sqrt{P_{2}(x)}\right)^{2},
    \label{eq:Hellinger_dist_def}
\end{align}
which quantifies the distance between two probability distributions $P_1(x)$ and $P_2(x)$.
Since $\Sigma=\mathcal{I}(S^{\prime},E^{\prime})+\mathcal{D}(\rho_{E}^{\prime}\|\gamma_{E})\ge\mathcal{D}(\rho_{E}^{\prime}\|\gamma_{E})$ from Eq.~\eqref{eq:quantum_EP_def_supp}, using Eqs.~\eqref{eq:Dalpha_monotonic} and \eqref{eq:fidelity_ineq}, we obtain
\begin{align}
    e^{-\Sigma}\le e^{-\mathcal{D}(\rho_{E}^{\prime}\|\gamma_{E})}\le\mathrm{Fid}(\rho_{E}^{\prime},\gamma_{E})\le\left[1-\mathrm{Hel}^{2}\left(P_{E}^{\prime},P_{E}\right)\right]^{2}.
    \label{eq:series_ineq_supp}
\end{align}
Let us define the mean and standard deviation of
the two general distributions $P_1(x)$ and $P_2(x)$ by $\chi_{i}\equiv\sum_{x}xP_{i}(x)$ and $\zeta_{i}\equiv\sqrt{\sum_{x}x^{2}P_{i}(x)-\chi_{i}^{2}}$ ($i = 1,2$), respectively. 
We employ the lower bound of the Hellinger distance proved in Ref.~\cite{Nishiyama:2020:HellingerBound}:
\begin{align}
    \mathrm{Hel}^{2}\left(P_{1},P_{2}\right)\geq1-\left[\left(\frac{\chi_{1}-\chi_{2}}{\zeta_{1}+\zeta_{2}}\right)^{2}+1\right]^{-\frac{1}{2}}.
    \label{eq:Hellinger_lower_bound}
\end{align}
Similar lower bounds of the Hellinger distance given means and variances can be found in Refs.~\cite{Holevo:1973:CRI,Dashti:2017:Bayes}, which are weaker than Eq.~\eqref{eq:Hellinger_lower_bound}. 
Combining Eqs.~\eqref{eq:series_ineq_supp} with \eqref{eq:Hellinger_lower_bound}, for an arbitrary measurement operator $G$ in the environment, we obtain
\begin{align}
    e^{-\Sigma}\le\left[\left(\frac{\mathbb{E}_{\rho_{E}^{\prime}}[G]-\mathbb{E}_{\gamma_{E}}[G]}{\sqrt{\mathrm{Var}_{\rho_{E}^{\prime}}[G]}+\sqrt{\mathrm{Var}_{\gamma_{E}}[G]}}\right)^{2}+1\right]^{-1}.
    \label{eq:F_moments_ineq}
\end{align}
Rearranging the terms in Eq.~\eqref{eq:F_moments_ineq}, we obtain
\begin{align}
    \left(\frac{\sqrt{\mathrm{Var}_{\rho_{E}^{\prime}}[G]}+\sqrt{\mathrm{Var}_{\gamma_{E}}[G]}}{\mathbb{E}_{\rho_{E}^{\prime}}[G]-\mathbb{E}_{\gamma_{E}}[G]}\right)^{2}\ge\frac{1}{e^{\Sigma}-1},
    \label{eq:QTUR_Hellinger_supp}
\end{align}
which is Eq.~\QTURUHellinger{} in the main text. 

\section{Relation to other uncertainty relations}

\subsection{Joint unitary transformation formalism}

In the main text, we derive the fundamental precision limits for the system described by the joint unitary evolution applied to the system and environment. 
Here, we review quantum thermodynamic uncertainty relations and the concept of entropy production in the joint unitary model.

In the joint unitary model, we prepare a state consisting of 
the system $S$ and the environment $E$.
Given the initial state $\rho_{SE} = \rho_S \otimes \gamma_E$,
where $\gamma_E$ is the Gibbs state,
we apply a unitary $U$ to obtain
\begin{align}
    \rho_{SE}^{\prime}=U(\rho_{S}\otimes\gamma_{E})U^{\dagger}.
    \label{eq:rho_SE_prime_supp}
\end{align}
In Refs.~\cite{Esposito:2010:EntProd,Reed:2014:Landauer},
the entropy production is studied for the joint unitary model. 
We can identify the heat flowing from the reservoir as
\begin{align}
    \Delta Q=\mathrm{Tr}[H_{E}\gamma_{E}]-\mathrm{Tr}[H_{E}\rho_{E}^{\prime}],
    \label{eq:DeltaQ_supp}
\end{align}
where $H_E$ is the Hamiltonian of the environment and 
$\rho_E^\prime = \mathrm{Tr}_S[\rho_{SE}^\prime]$ is the density operator after the unitary. 
The entropy flow is defined as
\begin{align}
    \Delta\mathcal{S}_{e}=\beta\Delta Q,
    \label{eq:DeltaSe_def_supp}
\end{align}
where $\beta$ is the inverse temperature of the environment $E$. 
The change in entropy of the system is
\begin{align}
    \Delta\mathcal{S}=-\mathrm{Tr}[\rho_{S}^{\prime}\ln\rho_{S}^{\prime}]+\mathrm{Tr}[\rho_{S}\ln\rho_{S}].
    \label{eq:system_entropy_supp}
\end{align}
According to the standard thermodynamic description,
the entropy $\Delta \mathcal{S}$, entropy flow $\Delta \mathcal{S}_e$, and entropy production $\Sigma$ satisfy
\begin{align}
     \Sigma=\Delta\mathcal{S}-\Delta\mathcal{S}_{e}.
     \label{eq:entropy_production_supp}
\end{align}
Reference~\cite{Esposito:2010:EntProd,Reed:2014:Landauer} showed that $\Sigma$ is expressed as the quantum relative entropy
 [Eq.~\eqref{eq:QRE_def_supp}]:
\begin{align}
    \Sigma&=\mathcal{D}(\rho_{SE}^{\prime}\|\rho_{S}^{\prime}\otimes\gamma_{E})\nonumber\\&=\mathcal{I}(S^{\prime},E^{\prime})+\mathcal{D}(\rho_{E}^{\prime}\|\gamma_{E})\ge0,
    \label{eq:quantum_EP_def_supp}
\end{align}
where $\mathcal{I}(\cdot,\cdot)$ is the quantum mutual information:
\begin{align}
    \mathcal{I}(S^{\prime},E^{\prime})\equiv \mathcal{S}(\rho_{S}^{\prime})+\mathcal{S}(\rho_{E}^{\prime})-\mathcal{S}(\rho_{SE}^{\prime}).
    \label{eq:ISE_def_supp}
\end{align}
It is known that the quantum mutual information is always non-negative, $\mathcal{I}(S^\prime,E^\prime) \ge 0$.

Reference~\cite{Salazar:2024:QRelEntTUR} derived a quantum thermodynamic uncertainty relation for the joint unitary model. 
Let $f(x)=1 / \sinh ^2[g(x) / 2]$, where
$g(x)$ is the inverse of $h(x)\equiv x \tanh (x / 2)$.
For an arbitrary measurement operator $G$ in the environment,
from Ref.~\cite{Salazar:2024:QRelEntTUR},
it can be stated that
\begin{align}
    \frac{\mathrm{Var}_{\rho_{E}^{\prime}}[G]+\mathrm{Var}_{\gamma_{E}}[G]}{\frac{1}{2}\left(\mathbb{E}_{\rho_{E}^{\prime}}[G]-\mathbb{E}_{\gamma_{E}}[G]\right)^{2}}\ge f\left(\frac{\Sigma+\Sigma^{*}}{2}\right),
    \label{eq:Salazar_bound_reply}
\end{align}
where
$\mathbb{E}_{\rho}\left[G\right] = \mathrm{Tr}\left[\rho G\right]$ and $\mathrm{Var}_{\rho}\left[G\right] = \mathrm{Tr}\left[\rho G^{2}\right] - \mathrm{Tr}\left[\rho G\right]^{2}$ as defined in the main text.
In Eq.~\eqref{eq:Salazar_bound_reply},
$\Sigma^*$ is defined by
\begin{align}
\Sigma^{*}&\equiv\mathcal{D}(\rho_{S}^{\prime}\otimes\gamma_{E}\|\rho_{SE}^{\prime}).\label{eq:Sigma_star_def_reply}
\end{align}
In Eq.~\eqref{eq:Salazar_bound_reply},
$\Sigma^*$ does not correspond to the standard entropy production. 
Although Eq.~\eqref{eq:Salazar_bound_reply} holds for a general joint unitary dynamics, the downside is that it depends on $\Sigma^*$, which has less physical interpretation. 
A similar relation can be derived using the properties of the quantum relative entropy. We can show that the following relation holds (see Section~\ref{sec:QTUR_Hellinger}):
\begin{align}
    \left(\frac{\sqrt{\mathrm{Var}_{\rho_{E}^{\prime}}[G]}+\sqrt{\mathrm{Var}_{\gamma_{E}}[G]}}{\mathbb{E}_{\rho_{E}^{\prime}}[G]-\mathbb{E}_{\gamma_{E}}[G]}\right)^{2}\geq\frac{1}{e^{\Sigma}-1}.
    \label{eq:QTUR_Hellinger}
\end{align}
Equation~\eqref{eq:QTUR_Hellinger} is similar to Eq.~\eqref{eq:Salazar_bound_reply}; however, it should be emphasized that the right-hand side only includes $\Sigma$. 
What these equations have in common is that they do not simply use the ratio of variance to mean, but rather employ two statistical quantities, one before and after the unitary evolution.
When the dynamics are trivial (i.e., $U = \mathbb{I}$), the entropy production vanishes ($\Sigma = 0$). In this case, both sides of Eq.~\eqref{eq:QTUR_Hellinger} diverge to infinity. This differs from Eq.~\gammaEUconvUTUR{} in the main text, which remains valid even when there is no dynamics.

So far, we are concerned with uncertainty relations whose lower bounds comprise the quantum entropy production. 
Let us comment on the difference between an uncertainty relation derived in Ref.~\cite{Hasegawa:2020:TUROQS}, which also considered the joint unitary dynamics. 
Reference~\cite{Hasegawa:2020:TUROQS} considered the system $S$ and environment $E$, where the initial environmental state is pure. 
Suppose that the initial environmental state is $\ket{0}$ without loss of generality. 
Then the time evolution in the system $S$ is represented by a Kraus representation:
\begin{align}
    \rho_S^\prime = \sum_g V_g \rho_S V_g^\dagger.
    \label{eq:Kraus}
\end{align}
For a general observable $G$ whose smallest eigenvalue is $0$, 
Ref.~\cite{Hasegawa:2020:TUROQS} showed the following relation:
\begin{align}
    \frac{\mathrm{Var}_{\rho_{E}^{\prime}}[G]}{\mathbb{E}_{\rho_{E}^{\prime}}\left[G\right]^{2}}\ge\frac{1}{\mathrm{Tr}\left[(V_{0}^{\dagger}V_{0})^{-1}\rho_{S}\right]},
    \label{eq:general_open_QTUR}
\end{align}
where $V_0$ is associated with the measurement of $G=0$. 
Using the Neumann series expansion, as demonstrated in Ref.~\cite{Ishida:2024:QTURVerification}, we can express $\left(V_{0}^{\dagger}V_{0}\right)^{-1}=\sum_{n=0}^{\infty}\left(\mathbb{I}-V_{0}^{\dagger}V_{0}\right)^{n}$. Under the first-order approximation, this result shows that the bound derived in Ref.~\cite{Hasegawa:2020:TUROQS} is equivalent to the Petrov inequality with 
a specific parameter value.
Although the quantum thermodynamic uncertainty relation in Ref.~\cite{Hasegawa:2020:TUROQS} depends on the system dynamics, as evidenced by the dynamics-dependent Kraus operator $V_0$ in Eq.~\eqref{eq:general_open_QTUR}, the precision limit we derive in Eq.~\gammaEUconvUTUR{} is dynamics-independent.

\subsection{Continuous measurement formalism\label{sec:continuous_measurement}}

The continuous measurement formalism is frequently employed in studies of thermodynamic uncertainty relations. An overview of this approach is provided below.

First, we consider the single reservoir case. 
Let $\beta$ denote the inverse temperature of the reservoir. 
The Lindblad equation describes the time evolution of an open quantum system:
\begin{align}
    \dot{\rho}(t)=\mathcal{L}\left(\rho(t)\right)=-i\left[H,\rho(t)\right]+\mathcal{K}\left(\rho(t)\right),
    \label{eq:Lindblad_def}
\end{align}
where $\rho(t)$ is the density operator at time $t$ and $\mathcal{L}(\rho)$ is the Lindblad superoperator and $\mathcal{K}(\rho)$ is the dissipator:
\begin{align}
\mathcal{K}(\rho)\equiv\sum_{m=1,2}\left[L_{m}\rho L_{m}^{\dagger}-\frac{1}{2}\{L_{m}^{\dagger}L_{m},\rho\}\right].
    \label{eq:dissipator_D_supp}
\end{align}
Here, $H$ denotes the Hamiltonian and $L_m$ denotes the $m$th jump operator. 
For simplicity, we assume two jump operators in this example. The approach easily generalizes to cases with more than two jump operators.
$L_1$ and $L_2$ are the pair of operators such that
\begin{align}
    [H,L_{1}]=-\omega L_{1},\hspace*{1em}[H,L_{2}]=\omega L_{2},
    \label{eq:HL_commutation_supp}
\end{align}
where $\omega$ is the energy difference. 
Moreover, $L_1$ and $L_2$ satisfy the following
detailed balance
relation: 
\begin{align}
    L_{1}^{\dagger}=e^{\beta\omega/2}L_{2}.
    \label{eq:LDB_supp}
\end{align}
Equations~\eqref{eq:HL_commutation_supp} and \eqref{eq:LDB_supp} guarantee that the steady state of Eq.~\eqref{eq:Lindblad_def} is the Gibbs state:
\begin{align}
    \mathcal{L}(\gamma) = \mathcal{K}(\gamma) = 0,
    \label{eq:Lindblad_eq_supp}
\end{align}
where $\gamma$ is the Gibbs state:
\begin{align}
    \gamma\equiv\frac{e^{-\beta H}}{Z},\hspace*{1em}Z\equiv\mathrm{Tr}[e^{-\beta H}].
    \label{eq:Gibbs_state_def_supp}
\end{align}
We now consider entropy production in the Lindblad equation.
From the first law of thermodynamics, heat and work can be identified as
\begin{align}
    \frac{d\mathbb{E}\left[H\right]}{dt}=\frac{d}{dt}\mathrm{Tr}\left[H\rho(t)\right]=\mathrm{Tr}\left[\dot{H}\rho(t)\right]+\mathrm{Tr}\left[H\dot{\rho}(t)\right]=\dot{\tilde{W}}+\dot{\tilde{Q}},
    \label{eq:W_and_Q_supp}
\end{align}
where $\dot{\tilde{W}}$ and $\dot{\tilde{Q}}$ denote the exerted work and the heat flowing into the system, respectively, in the Markov process 
(the tilde is used to indicate that these quantities are associated with the Markov process).
Here, the heat is expressed as
\begin{align}
    \dot{\tilde{Q}}=\mathrm{Tr}\left[H\dot{\rho}(t)\right]=\mathrm{Tr}\left[H\mathcal{L}\left(\rho(t)\right)\right]=\mathrm{Tr}\left[H\mathcal{K}\left(\rho(t)\right)\right].
    \label{eq:heat_def}
\end{align}
Calculating the time derivative of the entropy $\mathcal{S}(\rho)$, from the standard thermodynamic description, we can identify the entropy production as
\begin{align}
    \dot{\tilde{\Sigma}}=\dot{\mathcal{S}}-\dot{\tilde{\mathcal{S}}}_{e},
    \label{eq:EP_Lindblad_supp}
\end{align}
where $\dot{\tilde{\mathcal{S}}}_{e}$ is the entropy flow:
\begin{align}
    \dot{\tilde{\mathcal{S}}}_{e}=\beta\dot{\tilde{Q}}=\beta\mathrm{Tr}[H\mathcal{K}(\rho)].
    \label{eq:entropy_flow_Lindblad_supp}
\end{align}
In the joint unitary description, the heat is identified as the energy difference in the environment $E$ [Eq.~\eqref{eq:DeltaQ_supp}]. 
On the other hand, in the Lindblad formalism, the heat can be calculated from
quantities in the system alone. 
Note that the classical limit of $\tilde{\Sigma}$ is shown in Eq.~\eqref{eq:EP_sigma_def}. 

Next, we consider multiple reservoir cases,
where the inverse temperature of $\ell$th reservoir is given by $\beta_\ell$.
Suppose that there are $N_R$ reservoirs. 
When there are multiple reservoirs, we can decompose the dynamics
into those arising from each reservoir
\cite{Kosloff:2013:QThermoReview,VandenBroeck:2015:Review}:
\begin{align}
    \dot{\rho}(t)=\mathcal{L}\left(\rho(t)\right)=-i\left[H,\rho(t)\right]+\sum_{\ell=1}^{N_{R}}\mathcal{K}_{\ell}\left(\rho(t)\right),
    \label{eq:Lindblad_def2}
\end{align}
where $\mathcal{K}_{\ell}(\bullet)$ is the dissipator which results from the $\ell$th reservoir.
As in the single reservoir case, we assume that there are two jump operators for each reservoir and the pair of operators $L_{\ell,1}$ and $L_{\ell,2}$ satisfy Eqs.~\eqref{eq:HL_commutation_supp} and \eqref{eq:LDB_supp}. Under this assumption, the local Gibbs state becomes the steady-state solution of each dissipator:
\begin{align}
    \mathcal{K}_{\ell}(\gamma_{\ell})=0,
    \label{eq:K_gamma_r_supp}
\end{align}
where $\gamma_\ell$ is the local Gibbs state corresponding to the $\ell$th reservoir:
\begin{align}
    \gamma_{\ell}=\frac{e^{-\beta_{\ell}H}}{Z_{\ell}},\hspace*{1em}Z_{\ell}\equiv\mathrm{Tr}[e^{-\beta_{\ell}H}].
    \label{eq:local_Gibbs_state}
\end{align}
The heat is given by $\dot{\tilde{Q}}=\sum_{\ell=1}^{N_{R}}\dot{\tilde{Q}}_{\ell}$,
where $\tilde{Q}_\ell$ is the heat contribution from the $\ell$th reservoir,
$\dot{\tilde{Q}}_{\ell}\equiv\mathrm{Tr}\left[H\mathcal{K}_{\ell}(\rho)\right]$. 
The entropy flow in Eq.~\eqref{eq:EP_Lindblad_supp} can be identified as
\begin{align}
    \dot{\tilde{\mathcal{S}}}_{e}=\sum_{\ell=1}^{N_{R}}\beta_{\ell}\dot{\tilde{Q}}_{\ell}=\sum_{\ell=1}^{N_{R}}\beta_{\ell}\mathrm{Tr}[H\mathcal{K}_{\ell}(\rho)].
    \label{eq:entropy_flow_multi_supp}
\end{align}
The entropy production can be obtained from Eqs.~\eqref{eq:EP_Lindblad_supp} and \eqref{eq:entropy_flow_multi_supp}. 

The dynamical activity is another thermodynamic quantity, which is often employed in uncertainty relations.
The dynamical activity in the Lindblad equation is given by
\begin{align}
\dot{\tilde{A}}=\sum_{\ell=1}^{N_{R}}\mathrm{Tr}\left[L_{\ell,1}\rho L_{\ell,1}^{\dagger}+L_{\ell,2}\rho L_{\ell,2}^{\dagger}\right],
    \label{eq:DA_def_supp}
\end{align}
which quantifies the average number of jumps. 
Note that the classical definition of $\tilde{A}$ is shown in Eq.~\eqref{eq:DA_a_def}. 

Let $N_{\ell,1}$ denote the number of jump events induced by $L_{\ell, 1}$ in the $\ell$th reservoir during the time interval $[0,\tau]$. Similarly, $N_{\ell,2}$ is defined as the number of jump events induced by $L_{\ell, 2}$ in the same reservoir and time interval.
In the continuous measurement formalism, the observable is 
\begin{align}
    C=\sum_{\ell=1}^{N_{R}}\left(c_{\ell,1}N_{\ell,1}+c_{\ell,2}N_{\ell,2}\right).
    \label{eq:obs_C_def_supp}
\end{align}
Here, $c_{\ell,1}$ and $c_{\ell,2}$ are weights for the jumps.
When $c_{\ell,1}$ and $c_{\ell,2}$ satisfy the antisymmetry condition, $c_{\ell,1} = - c_{\ell,2}$, the observable corresponds to a thermodynamic current. 
For the classical limit of the Lindblad equation, the following relations are known to hold \cite{Barato:2015:UncRel,Gingrich:2016:TUP,Garrahan:2017:TUR,Terlizzi:2019:KUR}:
\begin{align}
    \frac{\mathrm{Var}[C]}{\mathbb{E}[C]^{2}}\ge\frac{2}{\tilde{\Sigma}},\,\,\,\frac{\mathrm{Var}[C]}{\mathbb{E}[C]^{2}}\ge\frac{1}{\tilde{A}}.
    \label{eq:TUR_KUR_supp}
\end{align}
Strictly speaking, the relation of the entropy production requires the antisymmetry condition, while the relation of the dynamical activity does not impose any constraint on the weights as long as $c_{\ell,1}$ and $c_{\ell,2}$ are real. 
In the presence of coherent dynamics, the expressions shown in Eq.~\eqref{eq:TUR_KUR_supp} do not hold. 
In quantum thermodynamic uncertainty relations based on dynamical activity, such as those discussed in Refs.~\cite{Hasegawa:2020:QTURPRL,Hasegawa:2023:BulkBoundaryBoundNC}, quantum coherence is included as part of the definition of dynamical activity.
A recent study \cite{Prech:2025:CoherenceQTUR} derived a thermodynamic uncertainty relation for dynamical activity in which the expectation value includes an additional term arising from coherent dynamics, rather than modifying the dynamical activity itself. 
Similarly, for quantum thermodynamic uncertainty relations based on entropy production, Refs.~\cite{Vu:2021:QTURPRL, Vu:2025:Fundamental} have shown that quantum coherence can enhance precision. This is achieved by introducing correction terms from coherent dynamics into the expectation value and the entropy production.

\section{Equality condition\label{sec:equality_condition}}

\subsection{Equality condition of Eq.~\gammaEUpnormUTUR{}}

We show the equality condition of the fundamental precision limit given by Eq.~\gammaEUpnormUTUR{}. 
The equality in Eq.~\gammaEUpnormUTUR{} requires that every inequality used in its derivation saturates. The main result is built from five inequalities; therefore, Eq.~\gammaEUpnormUTUR{} can be saturated only if all five hold with equality. We analyze the equality condition in reverse order.

First, Eq.~\PUminUeigenvalue{}:
\begin{align}
    \lambda_{\min}\!\left(\gamma_{E}\right)
    =\frac{e^{-\beta\epsilon_{\max}}}{\mathrm{Tr}\!\left[e^{-\beta H_{E}}\right]}
    \ge \frac{e^{-\beta\epsilon_{\max}}}{d_{E}e^{-\beta\epsilon_{\min}}}.
    \label{eq:lmin_ineq}
\end{align}
With $\Delta \epsilon=\epsilon_{\max}-\epsilon_{\min}>0$, the equality in Eq.~\eqref{eq:lmin_ineq} holds if $\beta=0$, in which case $\gamma_E$ is maximally mixed.

Next, Eq.~\evUsumUlowerbound{}:
\begin{align}
    \sum_{i=1}^{d_{S}}\lambda_{i}^{\uparrow}(\rho_{SE})
    \ge d_{S}\lambda_{\min}(\rho_{SE})
    = d_{S}\lambda_{\min}(\rho_{S})\lambda_{\min}(\gamma_{E}).
    \label{eq:sum_lambdai}
\end{align}
The equality holds when the $d_S$ smallest eigenvalues of $\rho_{SE}$ are all equal to $\lambda_{\min}(\rho_{SE})$. When $\beta=0$ (so $\gamma_E$ is uniform), this is ensured whenever $d_E\,\zeta \ge d_S$, where $\zeta$ is the degeneracy of $\lambda_{\min}(\rho_S)$. Therefore, the condition is automatically satisfied for $d_S \le d_E$.

For Eq.~\rOUexpr{}:
\begin{align}
    \lambda_{\min}(\rho_{E}^{\prime})
    \ge \sum_{i=1}^{d_{S}}\lambda_{i}^{\uparrow}(\rho_{SE}).
    \label{eq:lambdamin_rhoE}
\end{align}
Equality can be achieved by choosing $U$ to be a permutation (swap) unitary that rearranges the eigenvalues so that, after tracing out $S$, the sum of the $d_S$ smallest eigenvalues of $\rho_{SE}$ appears as the minimal diagonal entry of $\rho_E^\prime$.

The inequality used to pass from Eq.~\PetrovUineqIIUmain{} to Eq.~\pzeroUnormUTURUmethods{} is
\begin{align}
    \frac{1}{1-P(G=0)}\ge\frac{1}{1-\lambda_{\min}(\rho_{E}^{\prime})}.
    \label{eq:G0_lmin_ineq}
\end{align}
The equality holds when the eigenvector for $G=0$ matches the eigenvector of $\rho_E^\prime$ associated with $\lambda_{\min}(\rho_E^\prime)$.

Finally, for the Petrov inequality (Eq.~\PetrovUineqIIUmain{}),
\begin{align}
    \frac{\mathbb{E}_{\rho_{E}^{\prime}}\!\left[G^{s}\right]^{r/(s-r)}}
         {\mathbb{E}_{\rho_{E}^{\prime}}\!\left[G^{r}\right]^{s/(s-r)}}
    \ge\frac{1}{1-P(G=0)},
    \label{eq:E_PG0}
\end{align}
the equality is attained for a two-point distribution (see Section~\ref{sec:Petrov_equality}).
To summarize, the following conditions must be satisfied for the equality.
\begin{itemize}
\item $\beta=0$ ($\gamma_E$ is maximally mixed)
        \item $U$ is chosen as a permutation/swap that saturates Eq.~\eqref{eq:lambdamin_rhoE}
        \item The eigenvector corresponding to the minimum eigenvalue 0 of the observable $G$ agrees with the eigenvector corresponding to the minimum eigenvalue of $\rho_E^\prime$
        \item $G$ takes only two values
\item $d_E\,\zeta \ge d_S$ for the minimum-eigenvalue degeneracy $\zeta$ of $\rho_S$
\end{itemize}

\subsection{Equality condition of Petrov inequality\label{sec:Petrov_equality}}

Let us consider the equality condition of the Petrov inequality \cite{Valentin:2007:TailProb} for $b=0$. 
Suppose that $P(X)$ is a two-point distribution:
\begin{align}
    P(X=x_1) = p_1,P(X=x_2) = p_2 = 1 - p_1,
    \label{eq:PX_twopoint_def}
\end{align}
where $x_1 =0$ and $x_2 > x_1$. 
The inequality in question is given by
\begin{align}
1 - p_{1} \geq \frac{\mathbb{E}\left[X^{r}\right]^{s/(s-r)}}{\mathbb{E}\left[X^{s}\right]^{r/(s-r)}}.
\label{eq:Petrov_b0}
\end{align}
The expectation of $X^n$ is:
\begin{align}
\mathbb{E}[X^{n}]=x_{1}^{n}p_{1}+x_{2}^{n}p_{2}=x_{2}^{n}p_{2}.
\label{eq:EX_twopoint}
\end{align}
By using Eq.~\eqref{eq:EX_twopoint},
the right-hand side of Eq.~\eqref{eq:Petrov_b0} becomes
\begin{align}
    \frac{\mathbb{E}\left[X^{r}\right]^{s/(s-r)}}{\mathbb{E}\left[X^{s}\right]^{r/(s-r)}}=1-p_{1},
    \label{eq:Petrov_rhs}
\end{align}
which shows that the equality of Eq.~\eqref{eq:Petrov_b0} holds for a two-point distribution. 
This shows that $P(X)$ being a two-point distribution is a sufficient condition for the equality. 

Next, we show that $P(X)$ being a two-point distribution is the necessary and sufficient condition for the equality. 
Let us define a new random variable $Y = (X|X>0)$, which is defined under the condition $X>0$.
For example, when the realization of $X$ is $[1, 2, 0, 3, 0]$, its corresponding realization of $Y$ is $[1,2,3]$. 
The following relation holds:
\begin{align}
    \mathbb{E}\left[X^r\right]=\mathbb{E}\left[Y^r\right] P(X>0).
    \label{eq:EX_EY_supp}
\end{align}
From Eq.~\eqref{eq:Petrov_b0}, we obtain
\begin{align}
    P(X>0)\geq\frac{\mathbb{E}\left[X^{r}\right]^{s/(s-r)}}{\mathbb{E}\left[X^{s}\right]^{r/(s-r)}}=\frac{\mathbb{E}\left[Y^{r}\right]^{s/(s-r)}P(X>0)}{\mathbb{E}\left[Y^{s}\right]^{r/(s-r)}},
    \label{eq:PX0_ineq_supp}
\end{align}
which yields
\begin{align}
    \mathbb{E}\left[Y^{s}\right]\geq\mathbb{E}\left[Y^{r}\right]^{s/r}.
    \label{eq:EY_ineq_supp}
\end{align}
By Jensen's inequality, the equality in Eq. \eqref{eq:EY_ineq_supp} holds when $Y$ is constant, $Y = c > 0$. In terms of the original variable $X$, this corresponds to $X$ taking only the values $0$ and $c$, that is, a two-point distribution.

\section{Examples}

\begin{figure}
\includegraphics[width=0.85\linewidth]{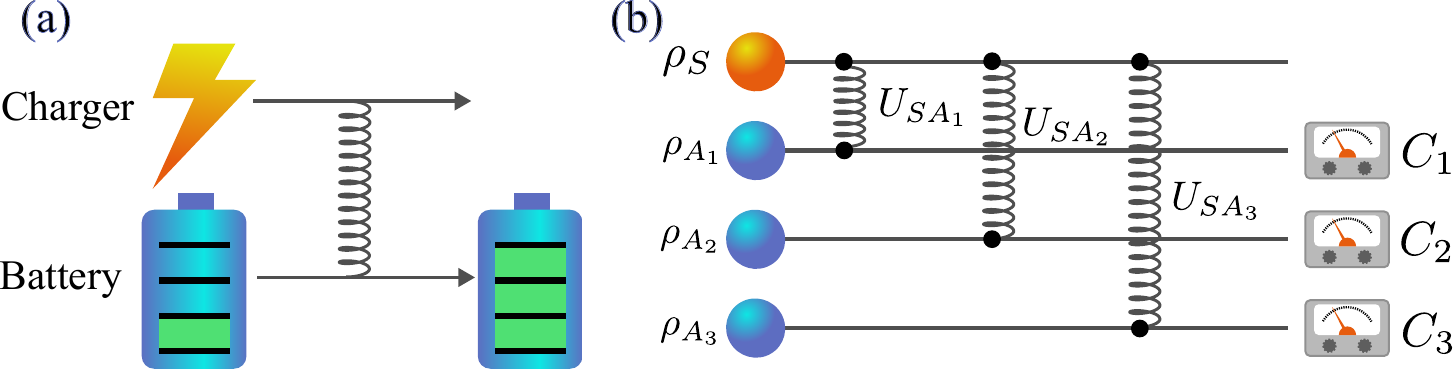}
\caption{
Applications of the joint unitary scenarios considered in the main text. 
(a) Charging of a quantum battery. The charger, which initially contains energy, is coupled to the battery. The battery is charged via the interaction.
(b) Collision model. The system interacts sequentially with ancillae $A_1, A_2, A_3$ and each ancilla is measured after its interaction. This entire procedure is equivalent to postponing all measurements and performing them simultaneously at the final time.
}
\label{fig:examples_illustration}
\end{figure}

As applications of the fundamental precision limit,
we consider two practical scenarios: the
quantum battery model (Fig.~\ref{fig:examples_illustration}(a)) and the collision model (Fig.~\ref{fig:examples_illustration}(b)). 

\subsection{Quantum battery}

A quantum battery is an experimental energy storage device that uses quantum effects to store and release energy \cite{Campaioli:2018:QuantumBatteries,Campaioli:2024:QuantumBatteries}. They have the potential to charge faster, deliver higher power, and operate more efficiently than their classical counterparts.
In the pioneering work \cite{Alicki:2013:QB}, the quantum battery is modeled as 
a finite-level system $\sum_{i=1}^{d_{E}}\epsilon_{i}\ket{\epsilon_{i}}\bra{\epsilon_{i}}$, 
which is identical to
the environmental Hamiltonian defined in Eq.~\HSHEUdef{} in the main text.
To charge or release energy in the quantum battery, several approaches exist.
The simplest approach considers a unitary $U_\mathrm{ch}$, which is applied directly to the battery. 
Another option is the charger-mediated protocol, in which the charger is introduced to charge the battery. 
The charger initially contains energy which is transferred to the battery via the interaction (Fig.~\ref{fig:examples_illustration}(a)). 
When we identify the charger and the battery as the system $S$ and the environment $E$, respectively, the charger-mediated protocol falls into the open quantum model considered in the main text (Fig.~\FIGmodelUdescription{}(a) in the main text). 

In the quantum battery, the primary quantity of interest is the amount of energy charged in the battery.
When the Hamiltonian of the battery is $H_E$, 
the stored energy is given by
\begin{align}
    \mathcal{E}_{B}\equiv\mathbb{E}_{\rho_{E}^{\prime}}[H_{E}]-\mathbb{E}_{\rho_{E}}[H_{E}],\label{eq:store_energy_def}
\end{align}
where $\rho_E^\prime$ and $\rho_E$ are the final and initial battery states, respectively. 
A primary objective is to
design a quantum battery that
maximizes $\mathcal{E}_B$,
since such a quantum battery can store more energy. 
In recent years, the charging precision has been emphasized as an alternative metric for quantum batteries
\cite{Friis:2018:PrecisionQB,Bakhshinezhad:2024:TradeoffsQB}.
Studies \cite{Friis:2018:PrecisionQB,Bakhshinezhad:2024:TradeoffsQB} used the post-charging energy variance as a measure of precision and sought a quantum battery that makes this variance as small as possible.
This alternative metric, the charging precision, is given by
\begin{align}
    \mathcal{P}_{B}&\equiv\mathrm{Var}_{\rho_{E}^{\prime}}[H_{E}].
    \label{eq:battery_precision}
\end{align}
The performance of quantum batteries can be evaluated by $\mathcal{E}_B$ and $\mathcal{P}_B$;
a quantum battery with larger $\mathcal{E}_B$ and smaller $\mathcal{P}_B$ is preferable. 
It is natural to consider the quantity $\mathcal{P}_B/\mathcal{E}_B^2$;
a quantum battery with smaller $\mathcal{P}_B/\mathcal{E}_B^2$ achieves the maximization of $\mathcal{E}_B$ and the minimization of $\mathcal{P}_B$. 
Without loss of generality, we set $\lambda_{\min}(H_E) = 0$. 
Using Eq.~\gammaEUconvUTUR{} in the main text, $\mathcal{P}_B/\mathcal{E}_B^2$ has the following lower bound:
\begin{align}
    \frac{\mathcal{P}_{B}}{\mathcal{E}_{B}^{2}}\ge\frac{\mathrm{Var}_{\rho_{E}^{\prime}}[H_{E}]}{\mathbb{E}_{\rho_{E}^{\prime}}[H_{E}]^{2}}\ge\left(\frac{d_{E}}{d_{S}}e^{\Phi(\beta,\Delta\epsilon,\lambda_{\min}(\rho_{S}))}-1\right)^{-1},
    \label{eq:battery_bound}
\end{align}
which is Eq.~\EPUtradeoff{} in the main text.
Equation~\eqref{eq:battery_bound} shows that the stored energy $\mathcal{E}_B$ and the charging precision $\mathcal{P}_B$ are in a trade-off relation.

\subsection{Collision model}

The collision model is a framework to study open quantum dynamics \cite{Ciccarello:2022:Collision}. 
In the collision model, the environment is represented as a sequence of ancillae, and the system interacts sequentially with each ancilla for a short time. 
The collision model is widely used to study the thermodynamic aspects of quantum systems \cite{Scarani:2002:ThermalizingCollisionModel,Santos:2020:JFT,Gross:2018:ContMeasQubit,Rodrigues:2019:CoherentCollision}. Moreover, 
the collision model forms the basis for the continuous measurement formalism \cite{Landi:2023:CurFlucReviewPRXQ}, which is often considered in quantum thermodynamic uncertainty relations (see Section~\ref{sec:continuous_measurement}).

We outline the standard collision model framework. 
Let $\mathfrak{N}$ be the number of ancillae. 
The system $S$ interacts with a sequence of ancillae $A_1,A_2,\cdots,A_\mathfrak{N}$, each identically prepared in the same state $\rho_{A_n}=\rho_A$ (Fig.~\ref{fig:examples_illustration}(b)).
For simplicity, we assume that the ancilla is a qubit with the Hamiltonian:
\begin{align}
    H_A = \epsilon\ket{1}\bra{1},
    \label{eq:HA_def_supp}
\end{align}
where $\epsilon > 0$ is the energy gap between $\ket{0}$ and $\ket{1}$. 
Then, $\rho_A$ is expressed as the Gibbs state:
\begin{align}
    \rho_{A}=\frac{e^{-\beta H_{A}}}{\mathrm{Tr}[e^{-\beta H_{A}}]}=\frac{e^{-\beta\epsilon}}{1+e^{-\beta\epsilon}}\ket{1}\bra{1}+\frac{1}{1+e^{-\beta\epsilon}}\ket{0}\bra{0}.
    \label{eq:rhoA_def}
\end{align}
Each system-ancilla interaction has a duration $t$ and is implemented by the unitary $U_{SA_n}$. After interacting with $A_n$, the state of $S$, $\rho((n+1)t)$, is given by
\begin{align}
    \rho_{S}((n+1)t)=\mathrm{Tr}_{A_{n}}\left[U_{SA_{n}}\left(\rho_{S}(nt)\otimes\rho_{A_{n}}\right)U_{SA_{n}}^{\dagger}\right].
    \label{eq:collision_CPTP}
\end{align}
We apply measurements to each ancillary in each step. 
Suppose that we apply the following Hermitian operator to each ancilla:
\begin{align}
    C = 1\ket{1}\bra{1} + 0\ket{0}\bra{0}.
    \label{eq:C_def_supp}
\end{align}
This acts as the qubit number operator that indicates whether the ancilla is excited.
Let $m_n \in \{0,1\}$ be the outcome when we measure $A_n$ with $C$. 
Therefore, after interacting with $\mathfrak{N}$ ancillae,
we have a sequence of outputs $\bm{m}=[m_1,m_2,\cdots,m_\mathfrak{N}]$. 
In effect, this procedure is equivalent to postponing all measurements and performing them all at once at the end (Fig.~\ref{fig:examples_illustration}(b)).
When we identify all the ancillae as the environment $E$, 
the environment Hamiltonian is
\begin{align}
    H_{E}=\sum_{n=1}^{\mathfrak{N}}H_{A_{n}}.
    \label{eq:HE_HA_supp}
\end{align}
where
\begin{align}
    H_{A_{n}}\equiv\underbrace{\mathbb{I}\otimes\cdots\otimes\mathbb{I}}_{n-1}\otimes H_{A}\otimes\underbrace{\mathbb{I}\otimes\cdots\otimes\mathbb{I}}_{\mathfrak{N}-n}.
    \label{eq:HAn_def_supp}
\end{align}
The initial Gibbs state of $H_E$ is
\begin{align}
    e^{-\beta H_{E}}=e^{-\beta H_{A_{1}}}\otimes\cdots\otimes e^{-\beta H_{A_{\mathfrak{N}}}},
    \label{eq:ebetaHE_supp}
\end{align}
which is the product of each ancilla Gibbs state. 
Moreover, the unitary applied to the environment is given by
\begin{align}
    U=U_{SA_{\mathfrak{N}}}\cdots U_{SA_{2}}U_{SA_{1}}.
    \label{eq:U_USAN_USA1}
\end{align}
When we identify each measurement process as the joint measurement applied in the final state as shown in Fig.~\ref{fig:examples_illustration}(b), the joint measurement operator is given by
\begin{align}
    C_{\mathfrak{N}}\equiv\sum_{\bm{m}}g(\bm{m})\ket{\bm{m}}\bra{\bm{m}}.
    \label{eq:CN_def_supp}
\end{align}
Here, $g(\bm{m}) \ge 0$ is an arbitrary non-negative function of $\bm{m}$ as long as $g(\bm{0}) = 0$ is satisfied. 
This condition guarantees that the minimum eigenvalue of $C_\mathfrak{N}$ is $0$. 
For example, suppose that $g(\bm{m})$ is given by
\begin{align}
    g(\bm{m}) = \sum_{n=1}^\mathfrak{N} m_n.
    \label{eq:gm_example}
\end{align}
In this case, $C_\mathfrak{N}$ is the number operator that counts the number of $\ket{1}$ in the ancillae.

As noted above, the collision model forms the basis for the continuous measurement. 
Consequently, it is natural to compute the lower bound on the relative variance $\mathrm{Var}[g(\bm{m})]/\mathbb{E}[g(\bm{m})]^{2}$. 
Using the fundamental precision limit [Eq.~\gammaEUconvUTUR{}], we can compute the lower bound on the relative variance $\mathrm{Var}[g(\bm{m})]/\mathbb{E}[g(\bm{m})]^{2}$.
The dimension of the environment is $d_E = 2^\mathfrak{N}$. 
Moreover, the energy gap is
\begin{align}
    \Delta \epsilon = \lambda_{\max}(H_E) - \lambda_{\min}(H_E) = \mathfrak{N}\epsilon.
    \label{eq:energy_gap_collision_model}
\end{align}
Using these values, from Eq.~\gammaEUconvUTUR{}, we obtain
\begin{align}
    \frac{\mathrm{Var}[g(\bm{m})]}{\mathbb{E}\left[g(\bm{m})\right]^{2}}\ge\left(\frac{2^{\mathfrak{N}}}{d_{S}}e^{\Phi(\beta,\mathfrak{N}\epsilon,\lambda_{\min}(\rho_{S}))}-1\right)^{-1}.
    \label{eq:collision_model_TUR}
\end{align}
Equation~\eqref{eq:collision_model_TUR} is the precision limit in the collision model. Given that the dimension of the environment is exponential in $\mathfrak{N}$ and that the energy gap $\Delta \epsilon$ is proportional to $\mathfrak{N}$, the lower bound in Eq.~\eqref{eq:collision_model_TUR} is very loose.

In continuous measurements, the lower bound of the thermodynamic uncertainty relation is proportional to the inverse of the total time $\tau$,
which contrasts with the exponential dependence on $\mathfrak{N}$ in Eq.~\eqref{eq:collision_model_TUR}. 
This is because, whereas in continuous time the state change achievable in each step is limited, the discrete-time dynamics allows arbitrary unitary transformations.
This feature is also observed in the thermodynamic uncertainty relation for classical Markov processes.
Indeed, the lower bound of the discrete-time thermodynamic uncertainty relation is known to decrease exponentially with the number of steps \cite{Proesmans:2017:TUR,Hasegawa:2019:FTUR}.
This is because discrete-time Markov processes permit state changes that are not allowed in continuous-time dynamics.

\section{Different initial settings\label{sec:other_cases}}

\begin{figure}
\includegraphics[width=0.6\linewidth]{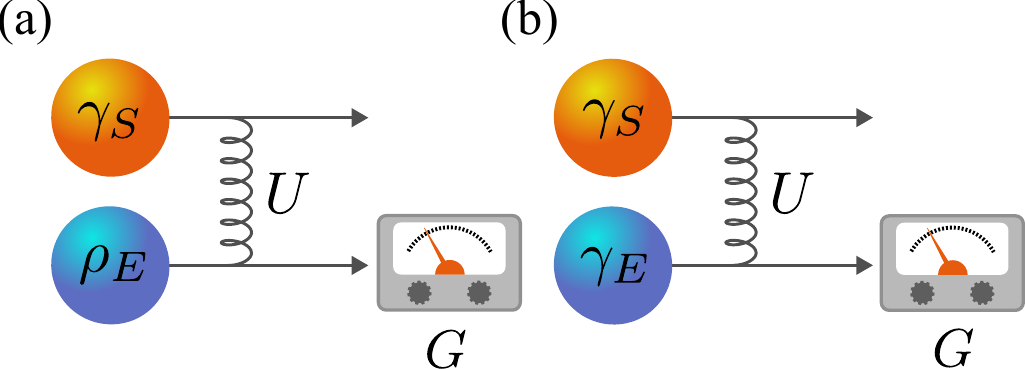}    
\caption{The joint unitary dynamics. The principal system $S$ and the environment $E$ undergo the joint unitary operator $U$, after which the environment is measured by the Hermitian operator $G$. The initial states are (a) $\rho_S=\gamma_S$ and (b) $\rho_S=\gamma_S$ and $\rho_E=\gamma_E$. 
}
\label{fig:model_description_other}
\end{figure}

In the main text, the initial state of the environment was assumed to be the Gibbs state. It is possible to consider cases where the principal system is in the Gibbs state or where both the principal system and the environment are in Gibbs states.

Consider the case where the initial state of the principal system $S$ is the Gibbs state, $\rho_S = \gamma_S$ (Fig.~\ref{fig:model_description_other}(a)). 
In this case, $\lambda_{\min}(\gamma_S)$ can be bounded from below by
\begin{align}
    \lambda_{\min}\left(\gamma_{S}\right)=\frac{e^{-\beta\sigma_{\max}}}{\mathrm{Tr}\left[e^{-\beta H_{S}}\right]}\ge\frac{e^{-\beta\sigma_{\max}}}{d_{S}e^{-\beta\sigma_{\min}}}.
    \label{eq:lambda_rhoS_lowerbound}
\end{align}
Using Eq.~\eqref{eq:lambda_rhoS_lowerbound}, we can repeat the same argument.
Using the Petrov inequality, we have
\begin{align}
\frac{\mathbb{E}_{\rho_{E}^{\prime}}\left[G^{s}\right]^{r/(s-r)}}{\mathbb{E}_{\rho_{E}^{\prime}}\left[G^{r}\right]^{s/(s-r)}}\ge\frac{1}{1-e^{-\Psi(\beta,\Delta\sigma,\lambda_{\min}(\rho_{E}))}},
    \label{eq:gammaS_pnorm_TUR}
\end{align}
where 
\begin{align}
    \Psi(\beta,\Delta\sigma,\lambda_{\min}(\rho_{E}))\equiv-\ln\lambda_{\min}(\rho_{E})+\beta\Delta\sigma.
    \label{eq:Psi_upperbound_def}
\end{align}
Equation~\eqref{eq:Psi_upperbound_def} corresponds to $\Phi$ defined in Eq.~\costUPhiUdef{}. 
Again, the energy bandwidth $\Delta \sigma \equiv \sigma_{\max} - \sigma_{\min}$ of the system Hamiltonian $H_S$ plays an important role, but unlike the case of Eq.~\gammaEUpnormUTUR{}, the dimensions of the system and the environment, $d_S$ and $d_E$, are not included explicitly.
Setting $r=1$ and $s=2$, the bound for the relative variance is obtained:
\begin{align}
\frac{\mathrm{Var}_{\rho_{E}^{\prime}}\left[G\right]}{\mathbb{E}_{\rho_{E}^{\prime}}\left[G\right]^{2}}\ge\left(e^{\Psi(\beta,\Delta\sigma,\lambda_{\min}(\rho_{E}))}-1\right)^{-1}.
    \label{eq:gammaS_conv_TUR}
\end{align}
Again using $s=\infty$ and $r=1$, we obtain the expectation bound:
\begin{align}
    \mathbb{E}_{\rho_{E}^{\prime}}[G]\le\lambda_{\max}[G]\left(1-e^{-\Psi(\beta,\Delta\sigma,\lambda_{\min}(\rho_{E}))}\right).
    \label{eq:gammaS_expectation_bound}
\end{align}

Next, we consider the scenario where both the initial states of the system and the environment are Gibbs states, denoted as $\rho_S = \gamma_S$ and $\rho_E = \gamma_E$ (Fig.~\ref{fig:model_description_other}(b)).
Then we obtain
\begin{align}
    \frac{\mathbb{E}_{\rho_{E}^{\prime}}\left[G^{s}\right]^{r/(s-r)}}{\mathbb{E}_{\rho_{E}^{\prime}}\left[G^{r}\right]^{s/(s-r)}}\ge\frac{1}{1-e^{-\Omega(d_{E},\Delta\sigma,\Delta\epsilon)}},
    \label{eq:pnorm_QTUR_dsigma_depsilon}
\end{align}
where $\Omega(d_E,\Delta\sigma,\Delta\epsilon)$ is defined by
\begin{align}
    \Omega(d_{E},\Delta\sigma,\Delta\epsilon)=\ln d_{E}+\beta\left(\Delta\sigma+\Delta\epsilon\right).
    \label{eq:Omega_def}
\end{align}
Again, by setting $r=1$ and $s=2$, we obtain
\begin{align}
    \frac{\mathrm{Var}_{\rho_{E}^{\prime}}\left[G\right]}{\mathbb{E}_{\rho_{E}^{\prime}}\left[G\right]^{2}}\ge\frac{1}{e^{\Omega(d_{E},\Delta\sigma,\Delta\epsilon)}-1}.
    \label{eq:QTUR_dsigma_depsilon}
\end{align}
Setting $r=1$ and $s=\infty$, we obtain
\begin{align}
    \mathbb{E}_{\rho_{E}^{\prime}}[G]\le\lambda_{\max}[G]\left(1-e^{-\Omega(d_{E},\Delta\sigma,\Delta\epsilon)}\right).
    \label{eq:gammaS_gammaE_expectation_bound}
\end{align}

\section{Numerical simulation}

\begin{figure}
\includegraphics[width=0.8\linewidth]{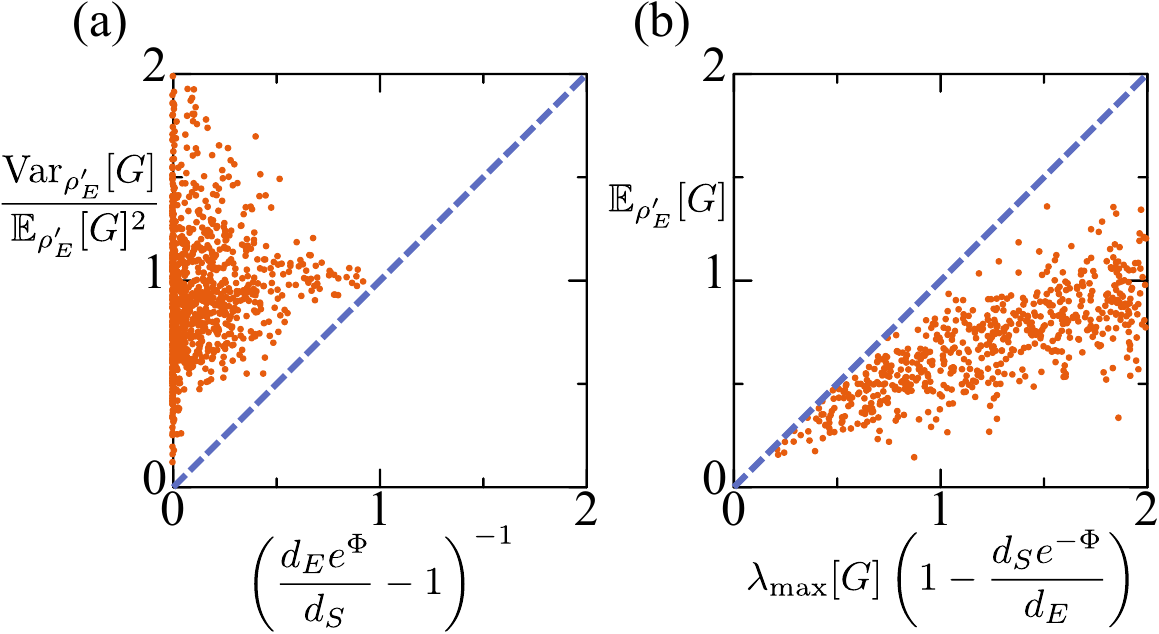}    \caption{Results of numerical simulation. 
(a)
Verification of Eq.~\gammaEUconvUTUR{}. $\mathrm{Var}_{\rho_{E}^{\prime}}[G]/\mathbb{E}_{\rho_{E}^{\prime}}[G]^{2}$ is plotted against $(d_E e^\Phi/d_S-1)^{-1}$. 
The points represent random realizations, while the dashed line indicates the equality case of Eq.~\gammaEUconvUTUR{}.
(b)
Verification of Eq.~\gammaEUexpectationUbound{}. $\mathbb{E}_{\rho_E^\prime}[G]$ is plotted against $\lambda_{\max}[G](1-d_S e^{-\Phi}/d_E)$. 
The points represent random realizations, while the dashed line indicates the equality case of Eq.~\gammaEUexpectationUbound{}.
In (a) and (b), the system dimension is set to $d_S = 2$. First, we randomly select the environment dimension $d_E$ from the set $\{2, 3, 4\}$. Then, we randomly generate the initial density matrix $\rho_S$ and a random Hamiltonian $H_E$. A randomly generated unitary $U$ is applied to the combined system.
The measurement operator $G$ is randomly generated, and we apply the operation $G \to G - \lambda_{\min}[G]\mathbb{I}$ so that the minimum eigenvalue becomes $0$. 
}
\label{fig:UTUR_GE}
\end{figure}

We perform numerical simulations to verify Eq.~\gammaEUconvUTUR{}. To do so, we randomly generate the density matrix of the system $\rho_S$ and the environmental Hamiltonian $H_E$, with the environment initially in the Gibbs state $\gamma_E$. We then apply a random unitary $U$ and a random Hermitian operator $G$ (with its minimum eigenvalue shifted to $0$). By repeating this process multiple times, we obtain random realizations for both sides of Eq.~\gammaEUconvUTUR{}: the left-hand side, $\operatorname{Var}_{\rho_E^{\prime}}[G] / \mathbb{E}_{\rho_E^{\prime}}[G]^2$, and the right-hand side, $\left(d_E e^{\Phi} / d_S-1\right)^{-1}$, which are plotted as points in Fig.~\ref{fig:UTUR_GE}(a).
The dashed line in Fig.~\ref{fig:UTUR_GE}(a) represents the equality condition of the bound. As we can see, all the points lie above this line, indicating that Eq.~\gammaEUconvUTUR{} holds true in our simulation. 
Similarly, in Fig.~\ref{fig:UTUR_GE}(b), we verify Eq.~\gammaEUexpectationUbound{} by plotting the left-hand side, $\mathbb{E}_{\rho_{E}^{\prime}}[G]$, against the right-hand side, $\lambda_{\max}[G]\left(1-d_{S}e^{-\Phi}/d_{E}\right)$. Once again, the dashed line in Fig.~\ref{fig:UTUR_GE}(b) represents the equality case of Eq.~\gammaEUexpectationUbound{}, and our numerical results confirm the validity of Eq.~\gammaEUexpectationUbound{}.